\documentclass[12pt,a4paper]{elsarticle}
\usepackage[top=25mm,bottom=25mm,left=2.4cm,right=2.4cm,asymmetric]{geometry}

\newcommand{\eref}[1]{(\ref{#1})}

\newcommand{\vect}[1]{\mathbf{#1}}
\newcommand{\vecsym}[1]{\boldsymbol{#1}}
\newcommand{\mtrx}[1]{\mathds{#1}}

\newcommand{\eeqn}{\end{equation}}

\usepackage{chngcntr}
\usepackage{amssymb}
\usepackage{amsmath}
\usepackage{bbm}
\usepackage{bm}
\usepackage{dsfont}
\usepackage[nodots]{numcompress}
\usepackage{url}
\usepackage{natbib}
\usepackage{units}
\usepackage{graphicx}
\usepackage{subfigure}
\usepackage{caption}
\usepackage{color}
\usepackage[titletoc,toc,title]{appendix}

\let\today\relax
\journal{International Journal}
\makeatletter
\def\ps@pprintTitle{%
   \let\@oddhead\@empty
   \let\@evenhead\@empty
   \let\@oddfoot\@empty
   \let\@evenfoot\@oddfoot
}
\makeatother

\begin{document}


\bibliographystyle{model2-names}\biboptions{authoryear}

\begin{frontmatter}

\title{3D analysis of a strain gradient plasticity material reinforced by elastic particles}

	



\author{Mohammadali Asgharzadeh}
\author{Jonas Faleskog\corref{CORR}}

\address{Solid Mechanics, Department of Engineering Mechanics, KTH Royal Institute of Technology, 10044 Stockholm, Sweden}

\cortext[CORR]{Corresponding author. E-mail: faleskog@kth.se}

\begin{abstract}

\noindent A 3D unit cell model containing eight different spherical particles embedded in a homogeneous strain gradient plasticity (SGP) matrix material is presented. The interaction between particles and matrix is controlled by an interface model formulated within the higher order SGP theory used. Strengthening of the particle reinforced material is investigated in terms of the increase in macroscopic yield stress. The results are used to validate a closed form strengthening relation proposed by the authors, which suggests that the increase in macroscopic yield stress is governed by the interface strength times the total surface area of particles in the material volume. 

\end{abstract}


\end{frontmatter}

\section{Introduction} 

\noindent Precipitation hardening is an efficient method to increase the strength of commercial alloys (\cite{Gladman1999}). The underlying mechanism for this strengthening operates on the scale of spacing between particles in relation to dislocations as discovered by Orowan (\cite{Orowan48}). Particles over a certain size or incoherent particles force mobile dislocations to bow around them, and by annihilating at the two meeting ends leave particles with so called Orowan loops. As dislocations are main carriers of plastic deformation, particles will be trapped in an elastic zone and plastic deformation will develop outside but in the immediate vicinity of the particles. Thus, steep plastic strain gradients arise at the particle-matrix boundaries which necessitates the existence of geometric necessary dislocations (GND) that lead to local hardening (\cite{Ashby70}). Hence, the presence of particles within the matrix makes the microscopic state of deformation heterogeneous, and this means that size scale effects play a major role also for the macroscopic stress-strain response of the material. As a consequence, classic continuum theory cannot be used to study this phenomenon, as it incorporates no material length scale and predicts no size effect. 

Dislocation mechanics has traditionally been used to study and developing analytical models for precipitation hardening (\cite{Orowan48}, \cite{friedel64}, \cite{ashby1969}, \cite{Ardell85}, \cite{Reppich93}, \cite{Deschamps99}). Computer simulations has since the early study by \cite{Foreman66} evolved into the field discrete dislocation dynamics for analysis of plastic deformation on small scales (\cite{Giessen95}, \cite{Espinosa2005}), and also been utilized to investigate precipitation strengthening (\cite{monnet2015multiscale}, \cite{hu2021modeling}). Even though particles on the micron scale are to too small to be characterized by conventional plasticity theory, they are usually too large to be analysed using approaches mentioned above (\cite{Hutchinson2000}). To bridge the gap in scales, strain gradient plasticity theories may be used to mimic plastic deformation induced by both statistically stored and geometrically necessary dislocations, respectively, see recent overviews by \cite{Lubarda16} and \cite{voyiadjis2019strain}. Several studies of precipitation strengthening based on SGP theories show promise. For example, a power law hardening matrix reinforced by rigid particles is investigated in \cite{Fleck97}, the model by \cite{Gao99} is used to study particle size effects with an axisymmetric model by \cite{Xue02}, and the theory of \cite{Gudmundson04} is employed to study size effects in metal matrix composites by \cite{Azizi14}.

In a most recent study on precipitation strengthening, \cite{Asgharzadeh2021} use the SGP theory of \cite{Gudmundson04} with one material length parameter $\ell$, that includes higher order stress and moment like variables which are captured by additional balance equations and higher order boundary conditions. Within this theory, a particle/matrix interface model was formulated that can be adapted for the strengthening response coupled to accommodation of GNDs at the surface of a particle. 
They propose a relation for the increase in yield stress $\sigma_{\rm p}$, which was deduced from numerical results obtained from a large and systematic set of axisymmetric finite element calculations. The relation is given on the form
\begin{equation} \label{eqn:Fs_function}
\sigma_{\rm p} = g_{\rm l} f \sigma_0 +  \frac{g_{nl}}{(1-f)} \: \frac{1}{V} \sum_{i=1}^{N_{\rm p}} ( \psi_{\Gamma} S^{\Gamma} )_i.
\end{equation}
The second term in \eref{eqn:Fs_function} is associated with particle/matrix interfaces. The strengthening contribution from particle $i$ is given by its surface area $(S^{\Gamma})_i$ times its interface strength $(\psi_{\Gamma})_i$. The total contribution from $N_{\rm p}$ particles in a material volume $V$ is then obtained by summation over all the particles. The interface strength is defined as $\alpha \sigma_0 \ell$, where $\sigma_0$ is the initial yield stress of the matrix material and $0 \le \alpha \le 1$ is non-dimensional parameter defining the ability of particle/matrix interface to resist plastic strains. The first term in \eref{eqn:Fs_function} is proportional to the volume fraction of particles $f$, and represents the strengthening contribution in absence of an interface strength ($\psi_{\Gamma} \rightarrow 0$). Furthermore, $g_l$ and $g_{nl}$ are functions that accounts for a mismatch in elastic constants between particles and matrix. When elastic constants match, $g_l=0$ and $g_{nl}=1$. Note that the second term in \eref{eqn:Fs_function} brings out a length scale, which in the case of identical particles can be identified as $N_{\rm p} S^{\Gamma} /V = 3f/r $, where $r$ is the particle radius.

In this work, the model proposed in \cite{Asgharzadeh2021} will be further investigated by use of a fully three-dimensional micromechanical model containing eight particles, where both a variation in particle size and spacing can be introduced. The structure of this article is as follows: In section 2, the micromechanical model, constitutive SGP formulation, particle/matrix interface model, and statistical features of the micromechanical model are presented. Section 3 briefly describes the 3D numerical FEM implementation of the model (a more comprehensive description is presented in Appendix A). In section 4, general model features are reviewed and the influence on strengthening of variations in particle size, spacing, and interface strength are presented. The paper is concluded in Section 5.

\section{Problem definition}

Of interest is the increase in yield strength of materials reinforced by elastic particles. Focus is on so-called 'hard' particles around which dislocations form loops, i.e., Orowan mechanism as discussed above. A key parameter in this mechanism is the average distance between particles, as the increase in yield strength is inversely proportional to this distance. The relative influence of this distance is directly proportional to the material length scale $\ell$, introduced by modelling the matrix as a strain gradient plasticity material, where the theory proposed by \cite{Gudmundson04} will be used. The effects of pile-up of dislocations at a particle boundary is simulated by the interface in the 2D axisymmetric analysis performed in \cite{Asgharzadeh2021}, 3D analysis will be conducted here to further examine the validity of \eref{eqn:Fs_function}.

\subsection{Micro-mechanical 3D model and statistical measures of particle size and spacing distribution}  \label{sec2-1}

\begin{figure}[htb!]
\begin{center}
		\subfigure[]{
    \includegraphics[scale=0.55]{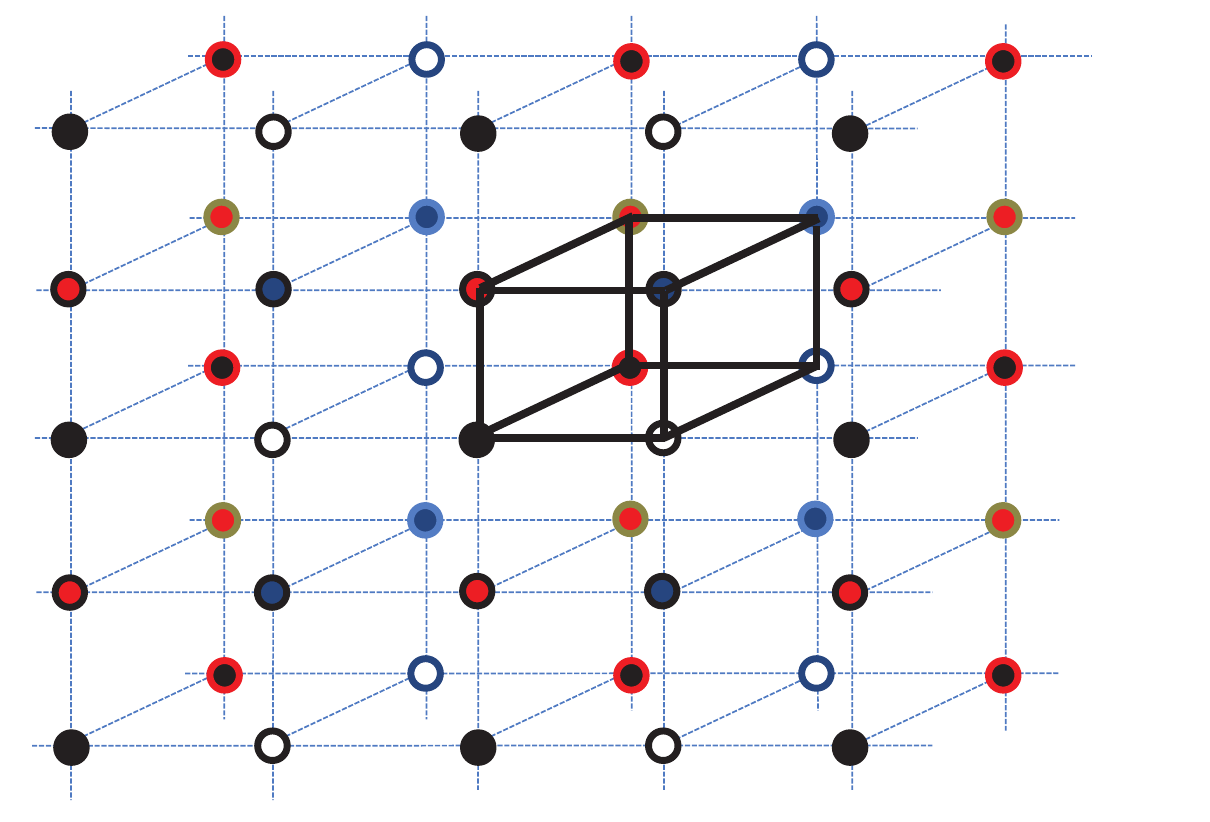}
		}
    \subfigure[]{
    \includegraphics[scale=0.25]{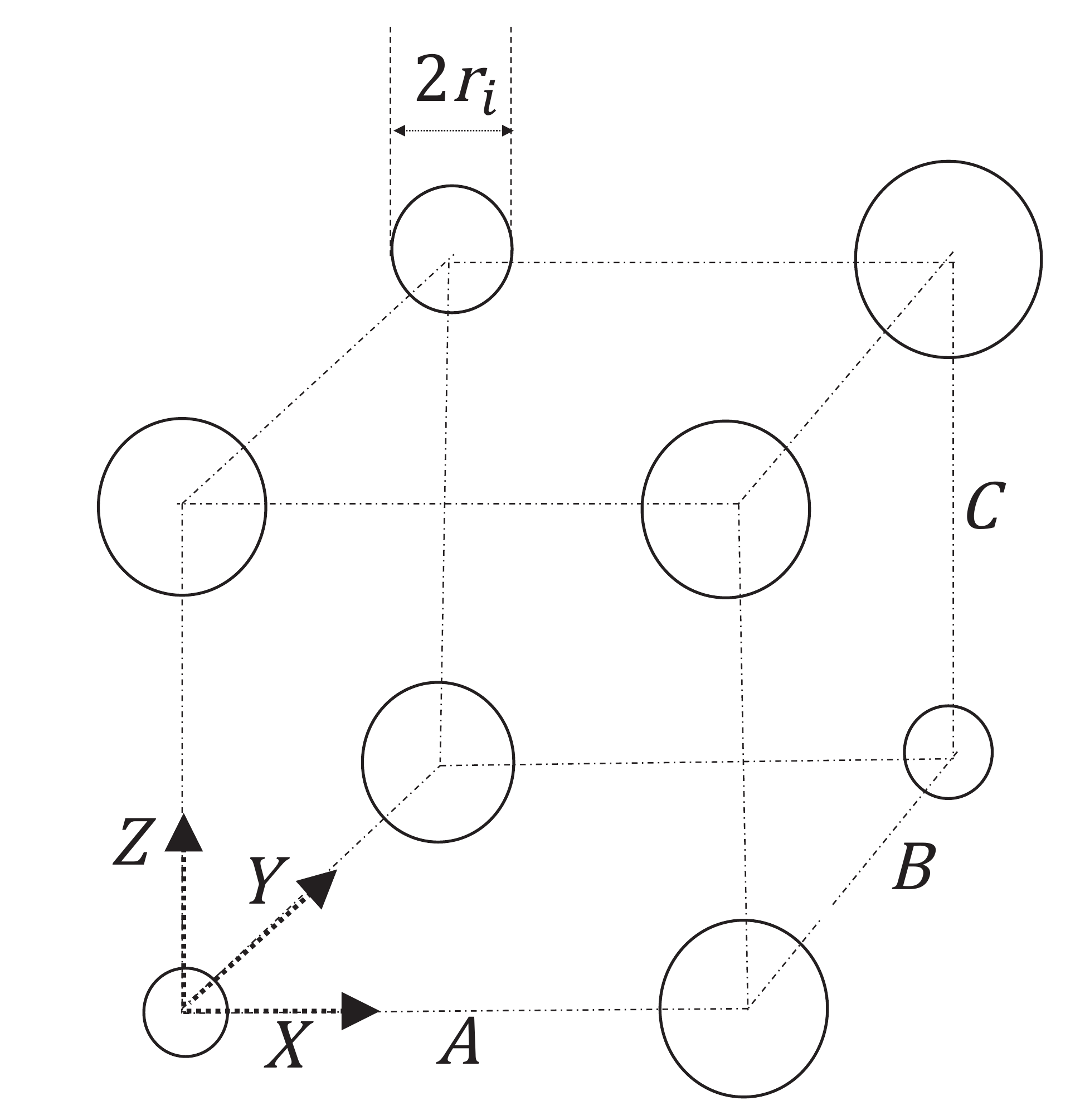}
    }
\caption{(a) The periodic distribution of particles within a matrix (only two particle layers into the depth are shown to avoid overcrowding). (b) Corresponding unit cell used in this study.}
\label{fig:Distribution_RVE}
\end{center}
\end{figure}

\noindent A periodic distribution of particles in a 3-dimensional space is considered as shown in Figure \ref{fig:Distribution_RVE}a. A unit cell containing eight different particles is identified from the periodic pattern, which is highlighted by the bold lines in this figure. The periodicity is obtained by 'mirroring' the unit cell in three directions and not by periodically 'repeating' it, so that every second layer in each direction is the same. Thus, the unit cell employed requires symmetric boundary conditions and not periodic.

The unit cell of cuboid shape has linear dimensions $A \times B \times C$ and is shown in Figure \ref{fig:Distribution_RVE}b. All particles are spherical with an individual radius of $r_i$, and only 1/8 of each particle is contained in the unit cell due to symmetry. The particles are modelled as linearly elastic with an interface layer that separates them from the matrix material that fills the space between them. The matrix material is modelled as a perfectly plastic strain gradient plasticity material.

Preliminary analysis revealed that a variation in ratio $B/A$ has exactly the same effect on the macroscopic response as a variation in ratio $C/A$. Therefore, $B=A$ and the influence of particle spacing will be addressed by a variation of $\xi = C/A$. Thus, the volume fraction of particles in the unit cell becomes
\begin{equation}\label{eqn:f_equation}
f = \frac{\dfrac{1}{8} \sum\limits_{i=1}^{8} \dfrac{4}{3} \pi r_i^{3}}{\xi A^3} = \{ r_i = r \} = \frac{4 \pi}{3 \xi} \left( \frac{r}{A} \right)^3,
\end{equation}
where the simplification in the last step is prompted by particles of equal size. The strengthening relation \eref{eqn:Fs_function} suggests that the increase in yield stress is proportional to the accumulated surface area times the interface strength of particles per volume, which for the 3D unit cell can be evaluated as
\begin{equation}
\frac{1}{V} \sum\limits_{i=1}^{8} ( \psi_{\Gamma} S^{\Gamma} )_i = \frac{\pi \sigma_0 \ell}{2\xi A^3} \sum\limits_{i=1}^{8} \alpha_{0i} r_i^2 = \{ r_i = r \} = (\frac{4 \pi}{3 \xi})^{1/3} ~ 3 f^{1/3}\sigma_0 \alpha_0 \dfrac{\ell}{A}, 
\end{equation}
where the last step results from identical particles of same size and use of \eref{eqn:f_equation}.

\begin{figure}[htb!]
\begin{center}
	\includegraphics[scale=0.75]{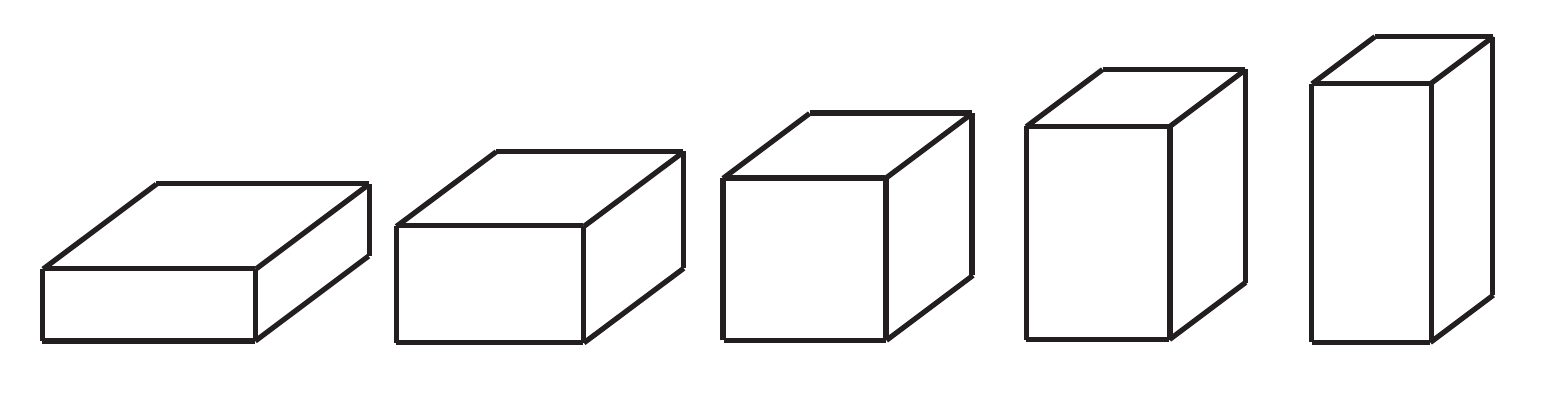}
\caption{Different unit cells resulting in various spacing distributions (from left to right: $C_{\rm v}^{\rm spacing}$ = 0.4, 0.2, 0.0, 0.2, 0.4).}
\label{fig:RVE_Sizes}
\end{center}
\end{figure}

Two measures will be used to describe the distribution: the average center-to-center distance between particles $L_{\rm p}$ and the coefficient of variance $C_{\rm v}^{\rm spacing}$. These measures will be evaluated by letting each particle be the center point of a 3D material unit defined by Voronoi tessellation. Such material unit will have 6 surface facets, which define the number of neighbours for each particle, which gives
\begin{equation}\label{eqn:Lp}
\begin{array}{lllllll}
&&& L_{\rm p} = A \cdot\theta(\xi) = \{r_i = r \} = \eta(\xi) r f^{-1/3}  & \qquad & \qquad & \vspace{2mm} \\
&&& S_{\rm D} = A \dfrac{\sqrt{2}}{3}|1-\xi|, \quad C_{\rm v}^{\rm spacing} = \dfrac{S_{\rm D}}{L_{\rm p}} = \sqrt{2} \dfrac{|1-\xi|}{(2+\xi)} , & \qquad & \qquad &  \vspace{2mm} \\
&&& \textrm{where} \quad \theta = \dfrac{2+\xi}{3}, \quad \eta = \dfrac{2+\xi}{3} (\dfrac{4\pi}{3\xi})^{1/3}. & \qquad & \qquad &
\end{array}
\end{equation}
A few examples of various unit cell shapes leading to different values of the coefficient of variance are depicted in Figure \ref{fig:RVE_Sizes}. As noted from \eref{eqn:Lp}, the same value of $C_{\rm v}^{\rm spacing}$ can be obtained for $\xi \lessgtr 0$. In Figure  \ref{fig:RVE_Sizes}, the values $\xi = C/A = [0.339 \quad 0.628 \quad 1.0 \quad 1.494 \quad 2.183]$ was used which gives $C_{\rm v}^{\rm spacing} = [0.4 \quad 0.2 \quad 0.0 \quad 0.2 \quad 0.4]$.

\begin{figure}[htb!]
	\begin{center}
		\subfigure[]{
			\includegraphics[scale=1]{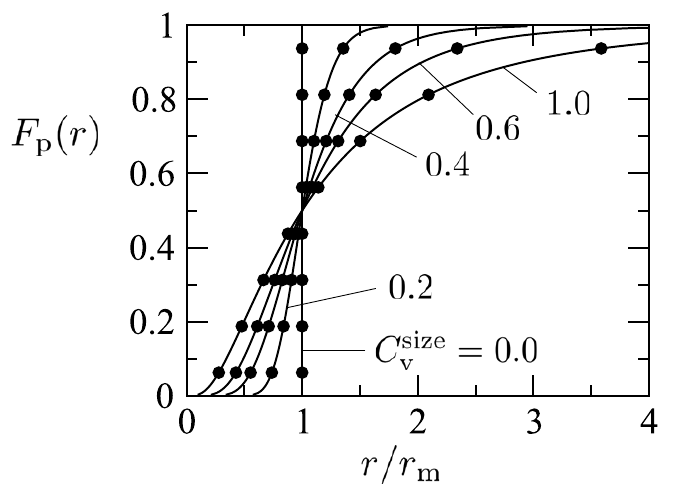}}
		\subfigure[]{
			\includegraphics[scale=0.75]{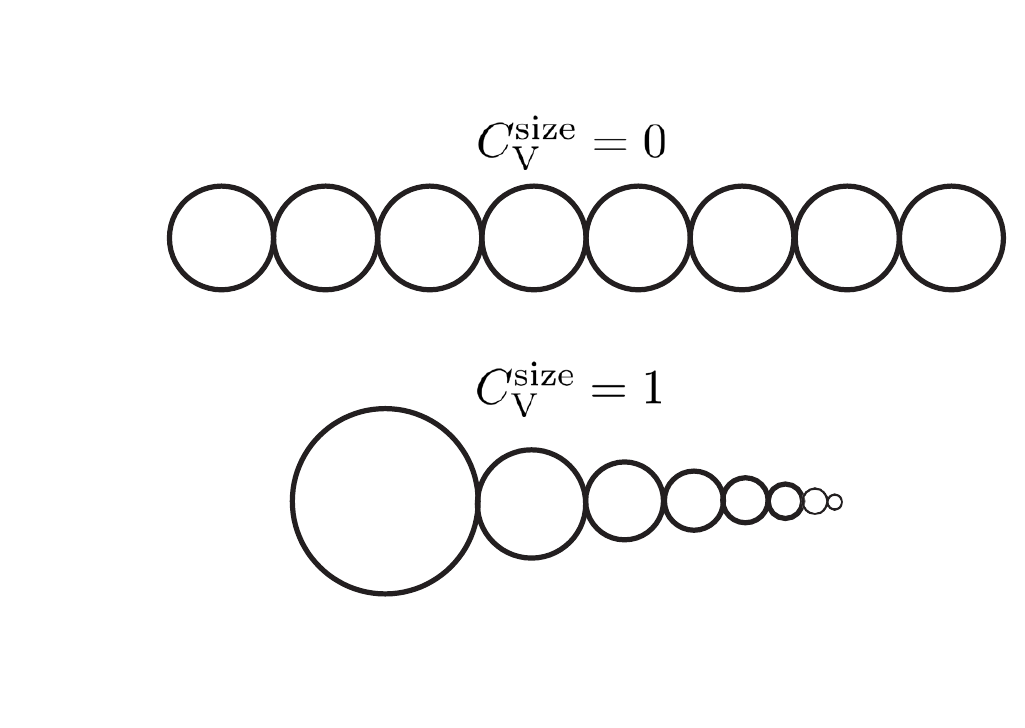}}
		\caption{Discrete particle sizes. (a) Five different sets of particle radii selected according to a log-normal distribution, where symbols represent the radius of each particle normalized by the median radius $r_{\rm m}$. (b) Two sets of discrete particle distributions with different $C_v$ scaled such that they correspond to the same volume fraction.}
		\label{fig:Particle_Size_distribution}
	\end{center}
\end{figure}

Variation in particle size will be introduced by aid of a log-normal distribution, described by the probability density function
\begin{equation}\label{eqn:pdf_D}
	f_{\rm{r}}(r;r_{\rm m},S_{\rm{N}}) =  \frac{1}{S_{\rm{N}} \sqrt{2\pi} r} \exp{\left[-\frac{1}{2}\left(\frac{\ln{(r/r_{\rm m})}}{S_{\rm{N}}}\right)^2\right]}.
\end{equation}
Here, $ r_{\rm m} $ is the median value (geometric mean) and $ S_{\rm{N}} $ is a dimensionless constant related to the standard deviation. The expected value (arithmetic mean), i.e., the mean particle radius is defined as
\begin{equation}\label{eqn:D0}
	r_{\rm 0} =  \int_0^{\infty} r f_{\rm{r}} {\rm{d}}r =  r_{\rm m} \exp(S_{\rm{N}}^2 / 2),
\end{equation}
and the standard deviation is given by
\begin{equation}\label{eqn:SD}
	S_{\rm D} =  \sqrt{\int_0^{\infty} (r - r_{\rm 0})^2 f_{\rm{r}} {\rm{d}}r}  = r_{\rm m} \exp(S_{\rm{N}}^2 / 2) \sqrt{\exp(S_{\rm{N}}^2) - 1}.
\end{equation}
The coefficient of variation of the size distribution is then given as
\begin{equation}\label{eqn:CV}
	c_{\rm V}^{\rm size} = S_{\rm D} / r_0 = \sqrt{\exp(S_{\rm{N}}^2) - 1}  ~ .
\end{equation}
The radius $r_i$ of particle $i$ is then determined by assuming an evenly distributed rank probability according to
\begin{equation}\label{eqn:pdf_ri}
	\int_0^{r_i} f_{\rm{r}}(r;r_{\rm m},S_{\rm{N}}) {\rm{d}}r = \frac{i-0.5}{8}
\end{equation}
Examples of the discrete particle size distribution resulting from \eref{eqn:pdf_ri} is illustrated in Figure \ref{fig:Particle_Size_distribution}a for $C_{\rm v}^{\rm size}$ in the range 0 to 1, where the discrete radii are represented as solid symbols. Figure \ref{fig:Particle_Size_distribution}b depicts two different selections of particle size distributions corresponding to $C_{\rm v}^{\rm size}$ equal to 0 and 1, i.e., the extreme cases shown in Figure \ref{fig:Particle_Size_distribution}a. The sizes of the particles are scaled such that both selections give the same overall volume fraction.

\subsection{SGP Theory}

\noindent The isotropic, higher order SGP theory proposed by \cite{Gudmundson04} was considered for the matrix material. The primary kinematic variables are displacements $u_{i}$ and plastic strains $\varepsilon_{ij}^{\rm p}$, related as
\begin{equation}\label{eqn:kinematic}
\varepsilon_{ij} = \frac{1}{2}(u_{i,j}+u_{j,i}), \quad \varepsilon_{ij} = \varepsilon_{ij}^{\rm e} + \varepsilon_{ij}^{\rm p},  \quad  \varepsilon_{kk}^{\rm p} = 0,
\end{equation}
Here, plastic incompressibility is assumed and $\varepsilon_{ij}^{\rm e}$ denote the elastic strains. In this higher order theory, plastic strains and their gradients, and the energy stored at internal boundaries (particle/matrix interfaces $S^{\Gamma}$) contribute to the internal work. In absence of body forces, balance between internal and external virtual work is expressed as
\begin{equation}\label{eqn:VirtualWork}
\int_{V}\left[ \sigma_{ij}\delta\varepsilon_{ij} + (q_{ij}-s_{ij})\delta\varepsilon_{ij}^{\rm p} + m_{ijk}\delta\varepsilon_{ij,k}^{\rm p} \right]{\rm d}V +  \int_{S^{\Gamma}}\left[ M^{\Gamma}_{ij}\delta\varepsilon^{\rm p}_{ij}  \right]{\rm d}S = \int_{S^{\rm ext}}\left[ T_{i}\delta u_{i} + M_{ij}\delta\varepsilon_{ij}^{\rm p} \right]{\rm d}S,
\end{equation}
where $ \sigma_{ij} $ is the Cauchy stress, $ s_{ij}$ is the deviatoric part of the Cauchy stress, $ q_{ij} $ is the micro stress, and $ m_{ijk} $ is the moment stress. On internal and external surfaces, the higher order moment tractions $M^{\Gamma}_{ij}$ and $M_{ij}$ arise, respectively. Equilibrium equations, natural boundary conditions, and conditions at the internal interface are obtained by integration by parts of \eref{eqn:VirtualWork} as
\begin{equation}\label{eqn:equilibrium}
\begin{array}{lcll}
  \sigma_{ij,j} = 0        & {\rm and} & m_{ijk,k} + s_{ij} - q_{ij} = 0 & \textrm{ in $V$}, \vspace{2mm} \\
  \sigma_{ij}n_{j} = T_{i} & {\rm and} & m_{ijk}n_{k} = M_{ij}           & \textrm{ on $ S^{\rm ext} $}, \vspace{2mm} \\
  & &  M^{\Gamma}_{ij} + m_{ijk}n^{\Gamma}_{k} = 0 & \textrm{ on $ S^{\Gamma} $}.
\end{array}
\end{equation}
Here, $n^{\Gamma}_{k}$ and $n_{k}$ denote the normal vectors on the internal and external surfaces, respectively.

The elastic behaviour is assumed to be isotropic with shear modulus $G$ and Poisson's ratio $\nu$ in both matrix and particles. The inelastic deformation in the matrix material is assumed to be purely dissipative, so that the rate of plastic dissipation can be expressed as
\begin{equation}\label{eqn:dissipation}
\dot{D} = q_{ij} \dot{\varepsilon}_{ij}^{\rm p} + m_{ijk}\dot{\varepsilon}_{ij,k}^{\rm p} = \Sigma \dot{E^{p}} \geq 0,
\end{equation}
where $\Sigma$ and $\dot{E^{p}}$ are work conjugate effective measures of stresses and strains. These effective measures, which reduce to standard $ J_{2} $-plasticity in absence of gradients effects, are defined as
\begin{equation}\label{eqn:EffectiveStress}
\Sigma = \sqrt{\frac{3}{2}\left( q_{ij}q_{ij} + \frac{m_{ijk}m_{ijk}}{\ell^{2}} \right)},
\end{equation}
\begin{equation}\label{eqn:EffectivePeeqRate}
\dot{E}^{\rm p} = \sqrt{\frac{2}{3}\left( \dot{\varepsilon}_{ij}^{\rm p}\dot{\varepsilon}_{ij}^{\rm p}   +  \ell^{2}\dot{\varepsilon}_{ij,k}^{\rm p}\dot{\varepsilon}_{ij,k}^{\rm p} \right) }.
\end{equation}
Here, $ \ell $ is a length parameter which sets the scale on which plastic strain gradients influence local hardening. Constitutive relations for inelastic strain rates that satisfy \eref{eqn:dissipation}-\eref{eqn:EffectivePeeqRate} is formulated as
\begin{equation}\label{eqn:const_qij}
\dot{\varepsilon}_{ij}^{\rm p} = \dot{\varepsilon}_{0} \frac{3q_{ij}}{2\Sigma}~\Phi(\Sigma,\sigma_{\rm f}),
\end{equation}
\begin{equation}\label{eqn:const_mijk}
\dot{\varepsilon}_{ij,k}^{\rm p} = \dot{\varepsilon}_{0} \frac{3m_{ijk}}{2\ell^{2}\Sigma}~\Phi(\Sigma,\sigma_{\rm f}),
\end{equation}
where $ \dot{\varepsilon}_{0} $ is a reference strain rate, and $ \Phi(\Sigma,\sigma_{\rm f}) =\dot E^p/\dot\varepsilon_0$ serve as a visco-plastic response function that depends on the effective stress $\Sigma$ and the inviscid flow stress $ \sigma_{\rm f} $. The matrix material is assumed to be perfectly plastic with yield stress $\sigma_0$ in the present study, hence $ \sigma_{\rm f} = \sigma_0$. The specific form of the visco-plastic response function used is (\cite{Dahlberg13a})
\begin{equation}\label{eqn:viscofcn}
\Phi(\Sigma,\sigma_{\rm f}) = \kappa\frac{\Sigma}{\sigma_{\rm f}} + \left(\frac{\Sigma}{\sigma_{\rm f}}\right)^{n}.
\end{equation}
This function of Ramberg-Osgood type remedies the indeterminacy of $q_{ij}$ and $m_{ijk}$ in the purely elastic regime (cf. \cite{Fredriksson07}), since a small but insignificant inelastic deformation will be present already upon initial loading. Nevertheless, the outcome of an analysis will reduce to a rate-independent elastic-plastic response, provided that a sufficiently small value of $\kappa$ and a high enough value of the exponent $n$ are chosen. In addition, the overall applied rate of loading should preferably be chosen such that the resulting plastic strain rate is of order $ \dot{\varepsilon}_{0} $ or less. Thus, a possible rate dependency in results can be avoided (cf. \cite{Anand05}), and effectively achieved by the choices $\kappa = 0.005 \varepsilon_{0}$ and $n = 2000$ (cf. \cite{Asgharzadeh2021}).

\subsection{Interface Theory} 

\noindent To account for GNDs piling up against particles in a phenomenological manner, the particle/matrix interface formulation suggested by \cite{Asgharzadeh2021} will be used as outlined below. By assuming that a surface energy $\Psi_{\Gamma}$ exists at the interface, the rate of dissipation at the interface can be expressed as
\begin{equation}\label{eqn:dissInt}
\dot{D}_{\Gamma } = \left( M^{\Gamma}_{ij} - \frac{\partial \Psi_{\Gamma}}{\partial \varepsilon_{ij}^{\rm p}} \right)\dot{\varepsilon}_{ij}^{\rm p} \geq 0.
\end{equation}
Here, it is assumed that $\dot{D}_{\Gamma } = 0$, such that the response of the interface is purely energetic. Moreover, the surface energy $\Psi_{\Gamma}$ is assumed to be a linear function of the effective plastic strain $\varepsilon_{\Gamma} = \sqrt{2 \varepsilon_{ij}^{\rm p} \varepsilon_{ij}^{\rm p} / 3}$, hence $\Psi_{\Gamma} = \Psi_{\Gamma}(\varepsilon_{\Gamma})$. It is convenient to introduce $ \psi_{\Gamma} = {\rm d} \Psi_{\Gamma} / {\rm d} \varepsilon_{\Gamma}$, thus the moment tractions at the interface and $ \psi_{\Gamma} $ are defined by
\begin{equation}\label{eqn:PsiInt}
M^{\Gamma}_{ij} = \frac{\partial \Psi_{\Gamma}}{\partial \varepsilon_{ij}^{\rm p}} =    \frac{{\rm d} \Psi_{\Gamma}}{{\rm d} \varepsilon_{\Gamma}} \frac{\partial \varepsilon_{\Gamma}}{\partial \varepsilon_{ij}^{\rm p}} =    \psi_{\Gamma} \frac{2}{3} \frac{\varepsilon_{ij}^{\rm p}}{\varepsilon_{\Gamma}} \quad \Rightarrow \quad \psi_{\Gamma} = \sqrt{\frac{3}{2} M^{\Gamma}_{ij}M^{\Gamma}_{ij} }
\end{equation}

Readers are referred to \cite{Asgharzadeh2021} for a discussion on a suitable form of $\psi_{\Gamma}(\epsilon_{\Gamma})$, which is given as 
\begin{equation}\label{eqn:FaleskogAsgharzadehFunction}
\psi_{\Gamma}(\epsilon_{\Gamma}) = \alpha_{0} \sigma_{0} \ell \cdot  \frac{ (\varepsilon_{\Gamma} / \kappa_{\Gamma})} {\left(1 + (\varepsilon_{\Gamma} / \kappa_{\Gamma})^{p}  \right)^{1/p}}  \thickspace .
\end{equation}
The first term represents the strength of an interface, where $\alpha_0$ is a non-dimensional parameter in the interval 0 to 1, which represents the ability of an interface to resist plastic flow. If $\alpha_0 = 0$ a micro-soft interface results and the non-local contribution to the proposed strengthening relation \eref{eqn:Fs_function} vanishes. By contrast, the interface will respond in a micro-hard manner if $\alpha_0 = 1$ and maximum contribution from the non-local term in \eref{eqn:Fs_function} is gained. The upper limit for $\alpha_0$ is proposed by \cite{Asgharzadeh2021} in a heuristic manner and this will be verified by 3D analyses presented below.

The second term in \eref{eqn:FaleskogAsgharzadehFunction} is introduced for numerical reasons, because the interface must have a finite initial "stiffness" in the implicit finite element implementation. Its effect on results is insignificant provided that $\kappa_{\Gamma} \ll \varepsilon_{0}$ and $p \gg 1$. Here, the values $\kappa_{\Gamma} = 0.1\varepsilon_{0}$ and $p = 5$ were used, which proved to be sufficient in this respect.

\section{Numerical implementation}

\begin{figure}[htb!]
\begin{center}
	\subfigure[]{
    \includegraphics[scale=0.65]{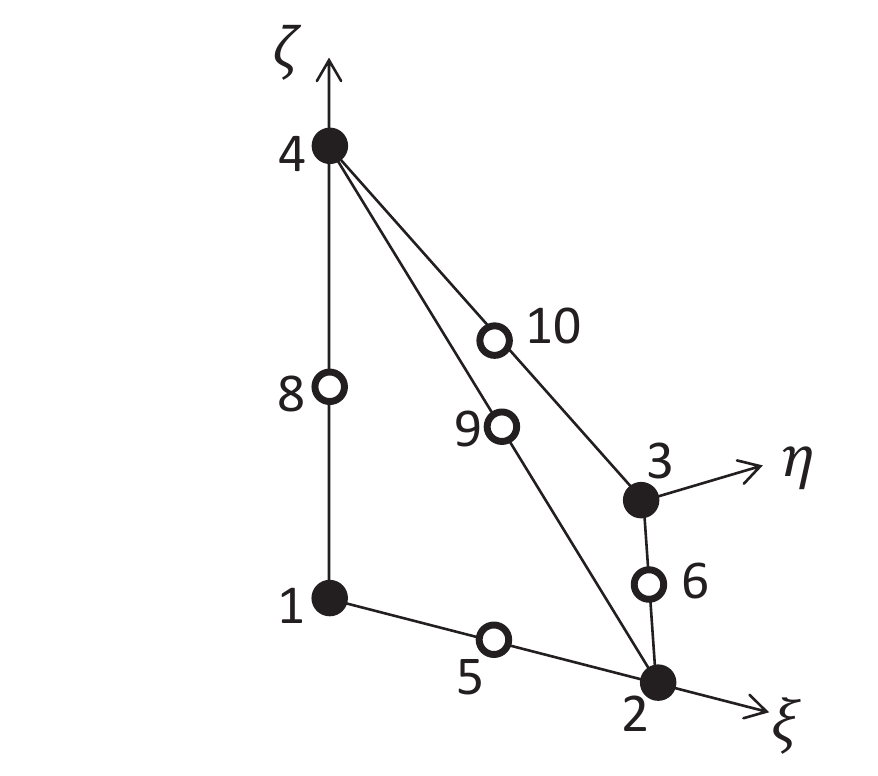}}
    \subfigure[]{
    \includegraphics[scale=0.65]{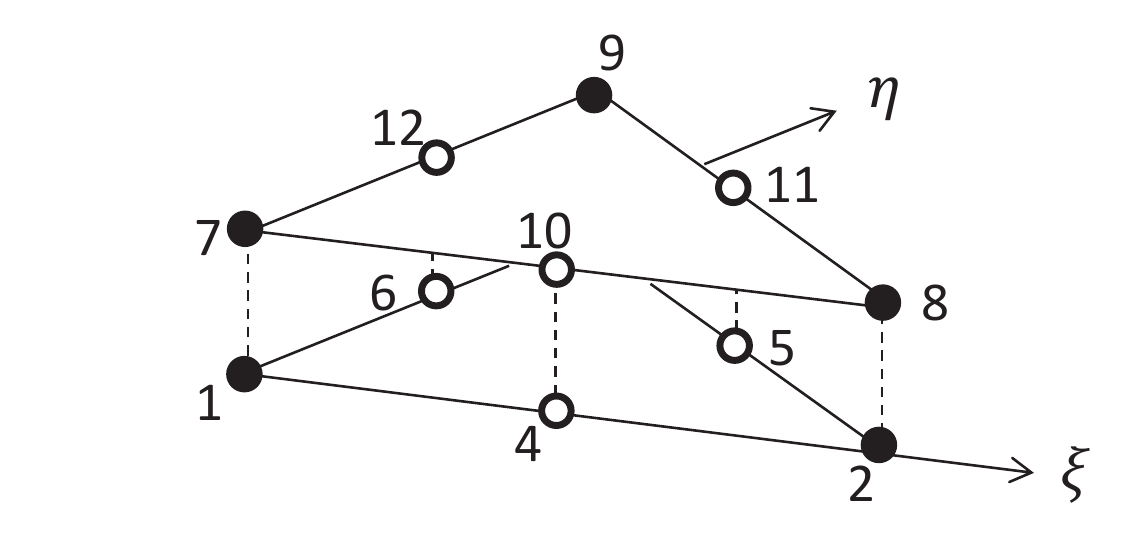}}
\caption{Elements used in this study. Solid circles denote nodes with plastic strain DOFs. (a) 3D continuum element, and (b) Interface element.}
\label{fig:Element_types}
\end{center}
\end{figure}

\noindent The micro-mechanical model introduced in Section \ref{sec2-1} was numerically solved by the finite element method. Three dimensional 10-node tetrahedron elements were used in particles and matrix, and 12-node interface (surface) elements were used to model the particle/matrix interaction. In both types of elements, quadratic interpolation is used for the displacement fields, and linear interpolation is used for the plastic strain fields. Hence, all nodes contain displacement degrees of freedom (DOFs), and only the vertex nodes contain plastic strain DOFs, as illustrated in Figure \ref{fig:Element_types}. The primary variables at the nodes in the 3D continuum element are described by the nodal displacement vector
\begin{equation}\label{eqn:dofu}
\vect{d}_{\rm u}^{\rm T} = [u_{x}^{1} \quad u_{y}^{1} \quad u_{z}^{1} \quad u_{x}^{2} \quad u_{y}^{2} \quad u_{z}^{2} \quad \ldots \quad u_{x}^{10} \quad u_{y}^{10} \quad u_{z}^{10}]_{30 \times 1},
\end{equation}
and the nodal plastic strain vector (only vertex nodes)
\begin{equation}\label{eqn:dofep}
\vect{d}_{\rm p}^{\rm T} = [{\epsilon_{xx}^{{\rm p}}}^{1} \quad {\epsilon_{yy}^{{\rm p}}}^{1} \quad {\gamma_{xy}^{{\rm p}}}^{1} \quad {\gamma_{xz}^{{\rm p}}}^{1} \quad {\gamma_{yz}^{{\rm p}}}^{1} \quad \ldots \quad {\epsilon_{xx}^{{\rm p}}}^{4} \quad {\epsilon_{yy}^{{\rm p}}}^{4} \quad {\gamma_{xy}^{{\rm p}}}^{4} \quad {\gamma_{xz}^{{\rm p}}}^{4} \quad {\gamma_{yz}^{{\rm p}}}^{4}]_{20 \times 1},
\end{equation}
Note that $ \gamma_{ij}^{\rm p} = 2\varepsilon_{ij}^{\rm p} $, and that the plastic strain in the $z$-direction is omitted due to plastic incompressibility ($ \varepsilon_{zz}^{{\rm p}} = - (\varepsilon_{xx}^{{\rm p}} + \varepsilon_{yy}^{{\rm p}})$). The DOFs for the primary variables in the interface element are described in a similar fashion. Details of the finite element discretization and implementation is presented in Appendix \ref{appendA}. This element technology is used in \cite{Fredriksson09}, \cite{Dahlberg13b}, and \cite{Asgharzadeh2021} to study 2D problems with the same higher order SGP theory.

The unit cell was subjected to uniaxial tension by applying a constant displacement rate $\dot{u}_{z} = \dot{u}_{z0} = {\rm constant}$ on boundary $ z = C $ (see Figure \ref{fig:Distribution_RVE}) with $u_z = 0$ at $z = 0$. Symmetry conditions were applied by $u_x = 0$ at $x = 0$ and $u_y = 0$ at $y = 0$ and enforced on boundaries $x = A $ and $Y = B$ by use of Lagrange multiplyers to keep these surfaces planar. The higher order boundary conditions are defined as
\begin{equation}
\begin{array}{ll}
M_{xx} = M_{yy} = M_{zz} = M_{yz} = 0, \thickspace \gamma_{xy}^{\rm p} = \gamma_{xz}^{\rm p} = 0  & {\rm on} \quad x = 0,~A  \\ \\
M_{xx} = M_{yy} = M_{zz} = M_{xz} = 0, \thickspace \gamma_{xy}^{\rm p} = \gamma_{yx}^{\rm p} = 0  & {\rm on} \quad y = 0,~B  \\ \\
M_{xx} = M_{yy} = M_{zz} = M_{xy} = 0, \thickspace \gamma_{xz}^{\rm p} = \gamma_{yz}^{\rm p} = 0  & {\rm on} \quad z = 0,~C  \\ \\
\varepsilon^{\rm p}_{ij} = 0  &  {\rm within \quad particles}.
\end{array}
\end{equation}
The last constraint is introduced to ensure that the particles will remain elastic during loading, which was achieved by prescribing $\vect{d}_{\rm p} = 0$ in elements belonging to a particle.

The elements described above are implemented into an SGP-FEM code, where a sparse solver (Pardiso) and OpenMP parallelization are utilized to speed up the computations. A fully backward Euler method is used to update stresses and to calculate a consistent tangent stiffness for the elements (cf. \cite{Asgharzadeh2021,Dahlberg2013b}).

\begin{figure}[htb!]
\begin{center}
	\includegraphics[width=1.0\textwidth]{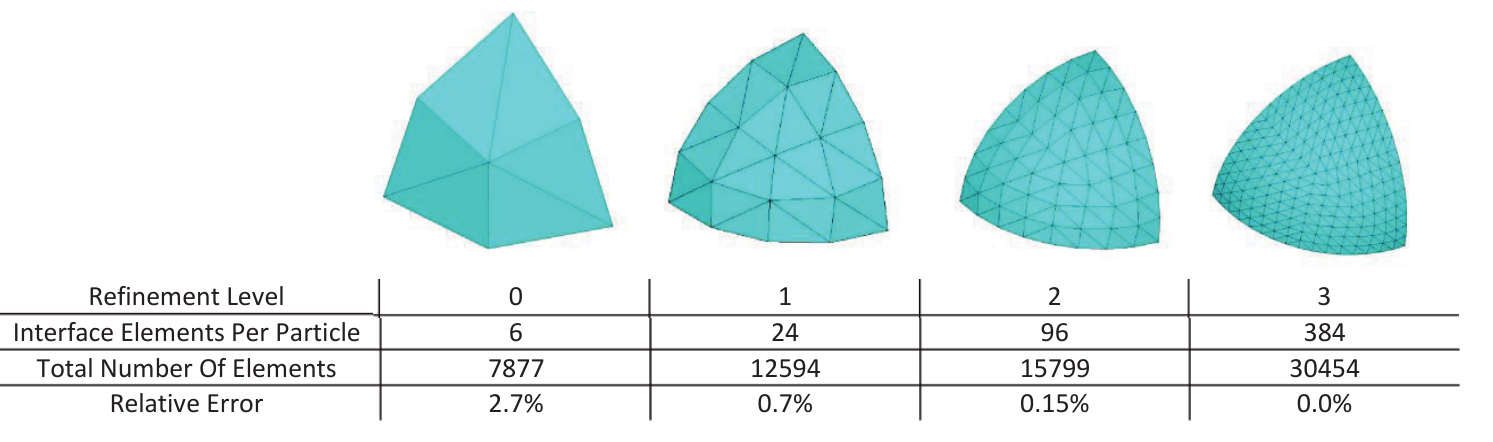}
\caption{Interface mesh densities used for the convergence study. Level 2 is chosen in all analyses in this work, except in those where $\alpha_0 \rightarrow 1$.}
\label{fig:convergence_study}
\end{center}
\end{figure}

\begin{figure}[htb!]
\begin{center}
		\subfigure[]{
    \includegraphics[scale=0.25]{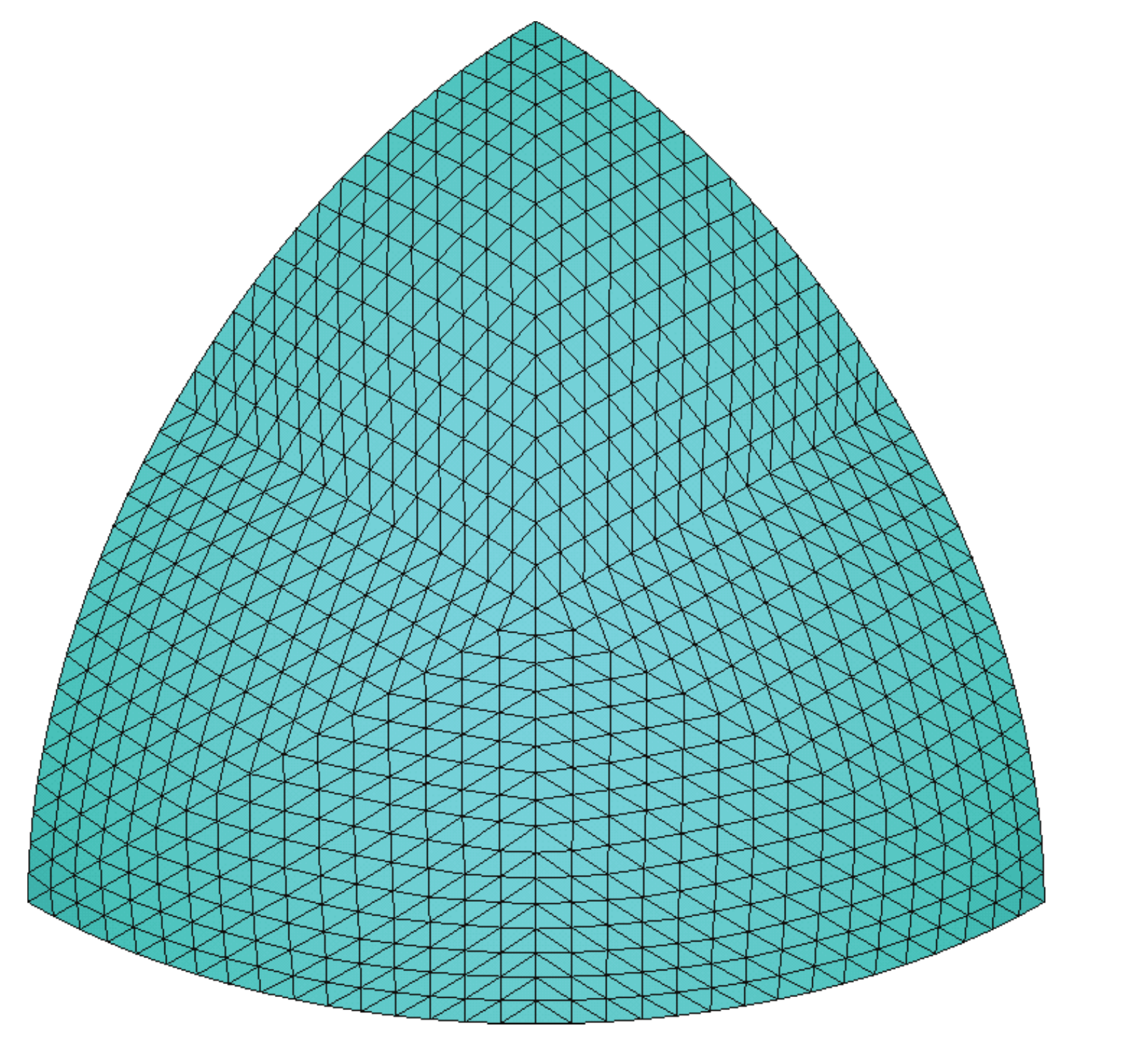}
		}
    \subfigure[]{
    \includegraphics[scale=0.24]{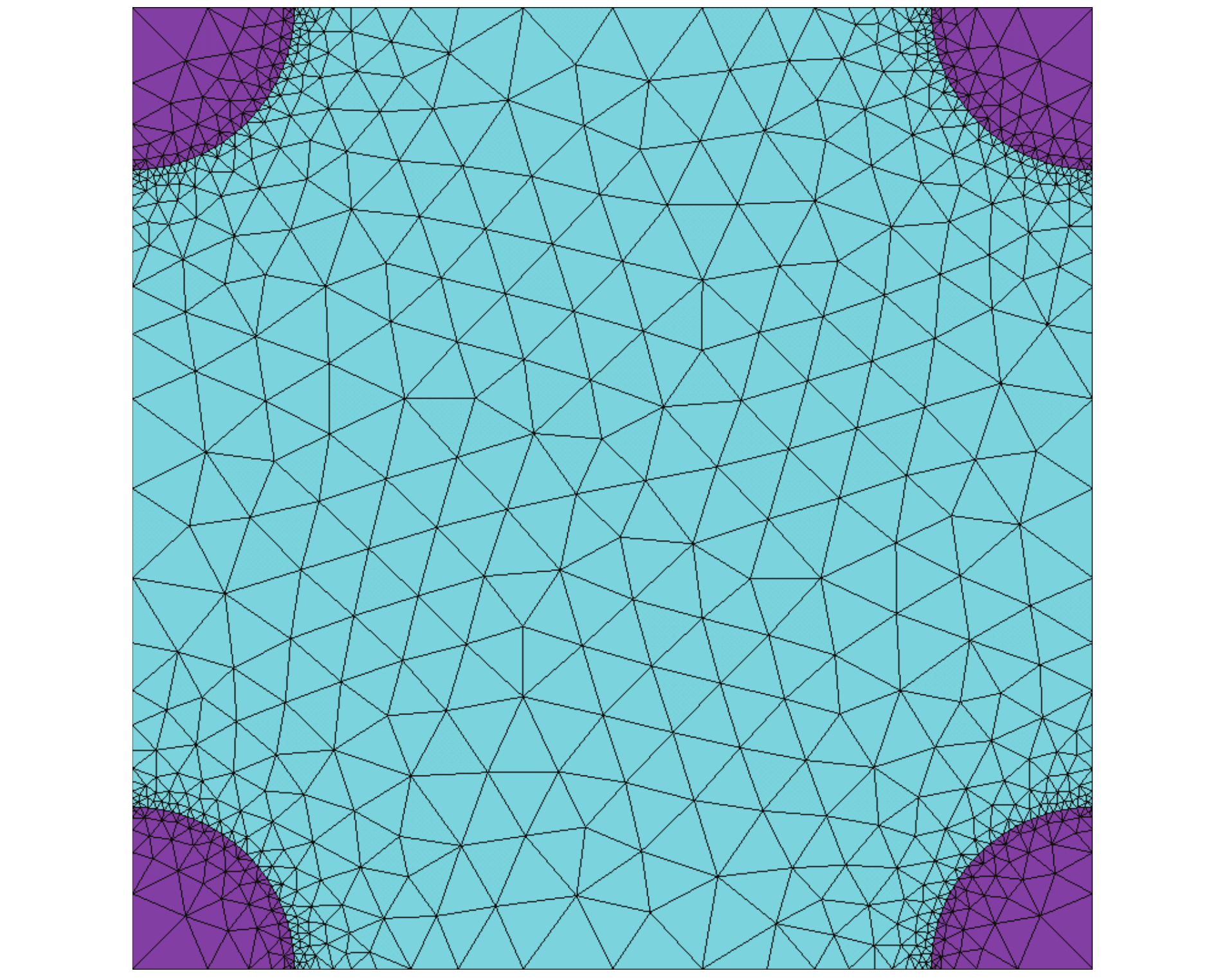}
    }
		\subfigure[]{
    \includegraphics[scale=0.32]{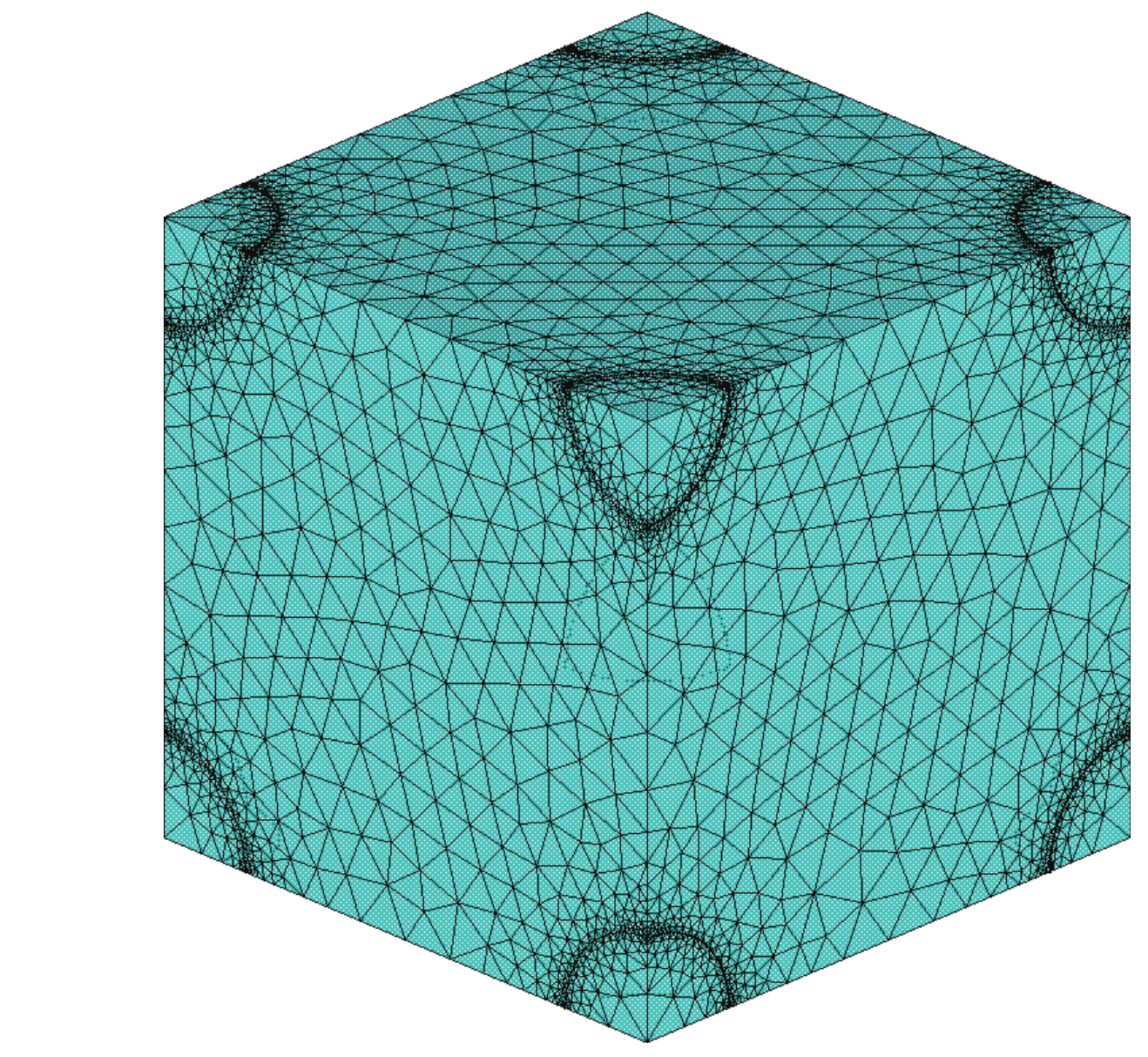}
    }
\caption{Most refined mesh employed for $\alpha_0 \rightarrow 1$. (a) Mesh density at the interface. (b) An outer boundary of the unit cell. (c) The full 3D mesh of the unit cell.}
\label{fig:fig_mesh}
\end{center}
\end{figure}

The accuracy of a numerical solution depends primarily on the resolution of the mesh at and near an interface. A convergence study was conducted with different mesh densities on the surface of a particle to establish a mesh that gives sufficiently accurate results. The study was conducted using $f=0.01$, $L_p/l = 0.464$ and $\alpha_0 = 0.49$ on four levels of mesh refinements as shown in Figure \ref{fig:convergence_study}. Accuracy was quantified based on the calculated yield stress, where the most refined mesh was used as a reference. It was judged that the second most refined mesh (level 2 in Fig. \ref{fig:convergence_study}) was adequate for most of the analysis performed in this study. However, for cases close to micro-hard interface conditions ($\alpha_0 \rightarrow 1$), where gradients of plastic strains are most pronounced, an even more refined mesh was employed. Figure \ref{fig:fig_mesh} show various images of this mesh.

Macroscopic stresses and strains were obtained as volume average values by homogenization. Homogenization involving higher order SGP theories is discussed in \cite{Dahlberg13b}, and for the boundary conditions employed in the current work, macroscopic strains is evaluated as
\begin{equation}\label{eqn:EijDefByU}
E_{ij} = \frac{1}{V}\int_{S^{\rm ext}} \frac{1}{2}\left( u_{i}n_{j} + u_{j}n_{i} \right)~{\rm d}S,
\end{equation}
and macroscopic stresses as
\begin{equation}\label{eqn:SigmaijDef}
\Sigma_{ij} = \frac{1}{V} \int_{S^{\rm ext}} \frac{1}{2} \left( \sigma_{ik}x_{j}n_{k} + \sigma_{jk}x_{i}n_{k} \right)  {\rm d}S,
\end{equation}
where $x_k$ are components of the position vector. The volume average of the effective deviatoric strain and effective stress can then readily be calculated as
\begin{equation}\label{eqn:EffStrainDef}
	E_{\rm e} = \sqrt{\frac{2}{3} E^{\rm d}_{ij}E^{\rm d}_{ij} } \quad \textrm{where } E^{\rm d}_{ij} = E_{ij} - \frac{1}{3}E_{kk}\delta_{ij},
\end{equation}
and
\begin{equation}\label{eqn:EffStressDef}
\Sigma_{\rm e} = \sqrt{\frac{3}{2} S_{ij}S_{ij}} \quad \textrm{where } S_{ij} = \Sigma_{ij} - \frac{1}{3}\Sigma_{kk}\delta_{ij}.
\end{equation}

\section{Results and discussion}

\noindent The strengthening relation \eref{eqn:Fs_function} rests on the 2D axisymmetric FEM calculations performed in \cite{Asgharzadeh2021}. Here, it will be scrutinized based on 3D finite element analysis of the cuboid unit cell model introduced above. First, general model features will be recaptured assuming equal sized particles in the 3D model. Then effects of particle spacing, and particle size variation will be investigated. Finally, the results from a few highly discriminating particle size and strength combinations chosen for validation of \eref{eqn:Fs_function} will be presented.

As the physical length scale of the 3D unit cell model is set by the matrix material length parameter $\ell$, characteristic dimensions of the micro-structure are related to $\ell$. The parameters subject to investigation are: $L_{\rm p}/ \ell$, $f$, $C_{\rm v}^{\rm size}$ (the inhomogeneity of particle size distribution), and $\xi=C/A$ (or alternatively the coefficient of variation of the spacing distribution $C_{\rm v}^{\rm spacing}$). The proposed relation for the increase in yield stress \eref{eqn:Fs_function} adapted to the current unit cell then becomes
\begin{equation} \label{eqn:Sp3Dcell_rvar}
	\sigma_{\rm p} / \sigma_0 = \frac{\pi}{2 \xi (1-f)} \: \frac{\ell}{A} \sum_{i=1}^{8} \left[ \alpha_{0i} \left( \frac{r_i}{A} \right)^2 \right],
\end{equation}
which for equal sized particles of radius $r$ that have the same interface strength simplifies to
\begin{equation} \label{eqn:Sp3Dcell_rconst}
	\sigma_{\rm p} / \sigma_0 = (2+\xi) \left(\frac{4 \pi}{3 \xi}\right)^{1/3} \frac{f^{2/3}}{(1-f)} \frac{\ell}{L_{\rm p}} \:  =  \: \frac{3 f}{1-f} \: \alpha_0 \: \frac{\ell}{r},
\end{equation}
where use of \eref{eqn:f_equation} gives the final step on the left-hand side.

\begin{figure}[htb!]
\begin{center}
  \includegraphics[scale=0.95]{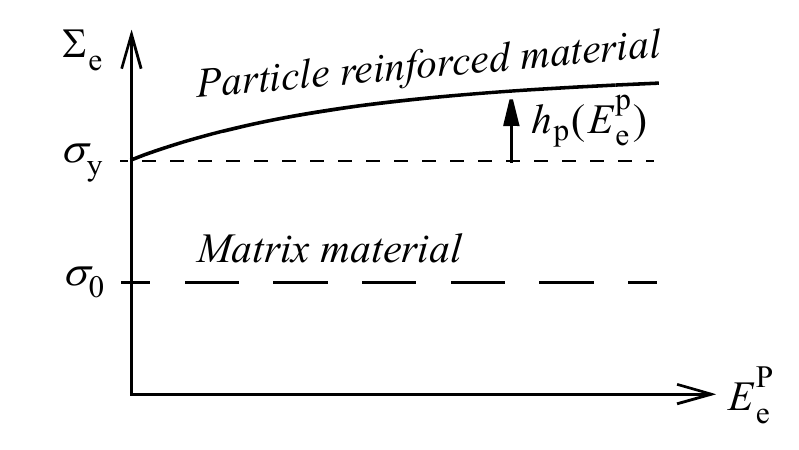}
  \caption{Definition of yield for pure matrix material (bottom dashed line) and precipitation-hardened material (top solid line).}
	\label{fig:Def_of_Sp}
\end{center}  
\end{figure}

All computations were carried out with Poisson's ratio $\nu = 0.3$ and shear modulus $G = \sigma_0 / [2\varepsilon_0(1+\nu)]$, with $\varepsilon_0 = 0.002$. The macroscopic initial yield stress for a particular case was determined by use of the $\Sigma_{\rm e}$ - $E^{\rm p}_{\rm e}$ curve extracted from the corresponding numerical solution, as schematically shown in Figure \ref{fig:Def_of_Sp}. Here, the macroscopic effective plastic strain $E^{\rm p}_{\rm e}$ was evaluated as $E^{\rm p}_{\rm e} = E_{\rm e} - \Sigma_{\rm e}/(3G)$. Note that the initial slope of the $\Sigma_{\rm e}$ - $E_{\rm e}$ response is equal to $3G$ for the choice of elastic constants considered in this work. To be specific, the macroscopic initial yield stress $\sigma_{\rm y}$, ideally identical to $\sigma_0 + \sigma_{\rm p}$, was determined from a linear approximation of the initial part of the $\Sigma_{\rm e}$ - $E^{\rm p}_{\rm e}$ curve (which essentially was linear) obtained from a linear regression analysis, which was evaluated at $E^{\rm p}_{\rm e} = 0$. In Figure \ref{fig:Def_of_Sp}, $h_{\rm p}(E_{\rm e}^{\rm p})$ signify the additional hardening caused by particles.

\subsection{General features and key results}

\begin{figure}[htb!]
\begin{center}
    \subfigure[]{ \includegraphics[scale=0.96]{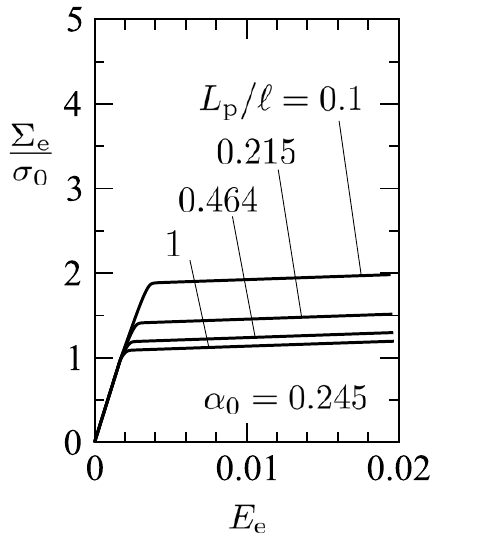} }
    \subfigure[]{ \includegraphics[scale=0.96]{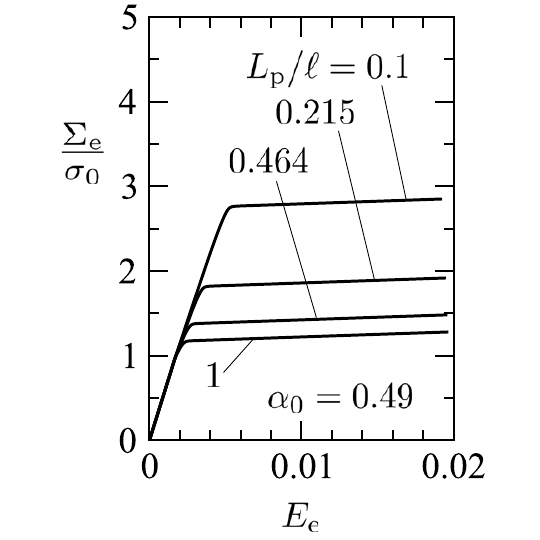} }
	\subfigure[]{ \includegraphics[scale=0.96]{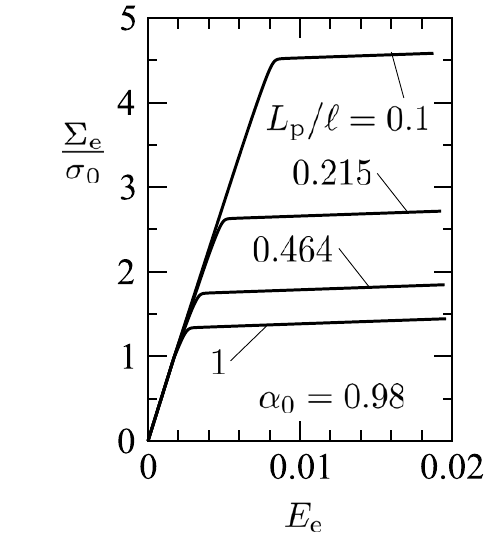} }
\caption{Macroscopic stress-strain curves for particles embedded in an elastic-perfect plastic matrix material ($N = 0.0$) for three different values of $\alpha_0$ = 0.245, 0.49, and 0.98 shown in graphs (a), (b) and (c), respectively ($f = 0.02$ for all cases).}
\label{fig:general_results}
\end{center}
\end{figure}

\noindent Figure \ref{fig:general_results} shows typical stress-strain curves resulting from the current model with a perfectly plastic matrix material. In graph (a), particles are furnished with a weak interface ($\alpha_0 = 0.245$), in graph (b) with medium strength interface ($\alpha_0 = 0.245$), and in graph (c) the interface are close to micro-hard ($\alpha_0 = 0.98$). Each graph contains four curves corresponding to four different values of center-to-center particle spacing $L_{\rm p} = \{0.1, 10^{-2/3}, 10^{-1/3}, 1\}$. All results in Figure \ref{fig:general_results} were generated with $\xi = 1$ and equal sized particles. The trend is clear, a decreasing particle spacing, and an increasing interface strength leads to a higher yield stress in agreement with  \eref{eqn:Sp3Dcell_rconst}.  

\begin{figure}[htb!]
\begin{center}
    \subfigure[]{ \includegraphics[scale=0.96]{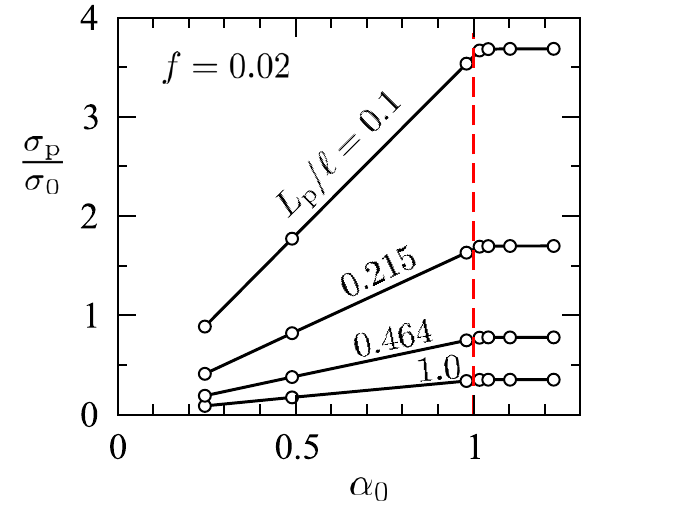} }
    \subfigure[]{ \includegraphics[scale=0.96]{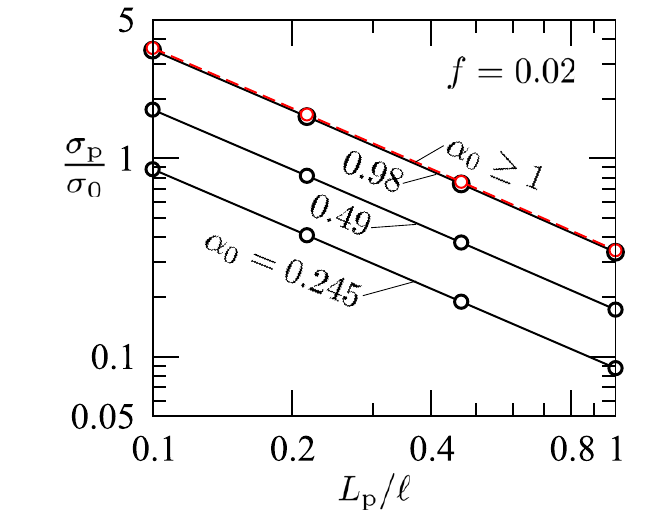} }
\caption{Influence of (a) interface strength and (b) particle spacing on the increase of yield strength.}
\label{fig:fig_key_results}
\end{center}
\end{figure}

The specific influence of parameters $\alpha_0$, $L_{\rm p}/\ell$ and $f$ on the increase in yield stress as surmised from Eq. \eref{eqn:Sp3Dcell_rconst} will now be examined in more detail. In Figure \ref{fig:fig_key_results} the isolated effects of $\alpha_0$ and $L_{\rm p}/\ell$ on $\sigma_{\rm p} / \sigma_0$ are brought out for $f = 0.02$. As clearly seen from Fig. \ref{fig:fig_key_results}(a), $\sigma_{\rm p} / \sigma_0$ is proportional to $\alpha_0$ until $\alpha_0 \approx 1$, where micro-hard conditions limit further strengthening, which applies independently of the $L_{\rm p}/\ell$ values considered. In Fig. \ref{fig:fig_key_results}(b), where $\sigma_{\rm p} / \sigma_0$ is plotted versus $L_{\rm p}/\ell$ in a log-log graph, $\sigma_{\rm p} / \sigma_0$ exhibits a distinct inverse dependency on $L_{\rm p}/\ell$, which is valid for $\alpha_0 \leq 1$. Again, confirming that micro-hard conditions limit further strengthening if $\alpha_0$ is chosen larger than one as shown by the red dashed line in Figure \ref{fig:fig_key_results}(b).

\begin{figure}[htb!]
\begin{center}
    \subfigure[]{ \includegraphics[scale=0.96]{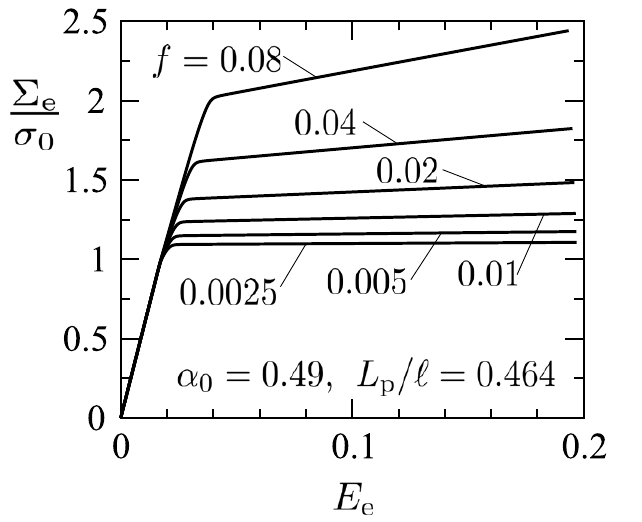} }
    \subfigure[]{ \includegraphics[scale=0.96]{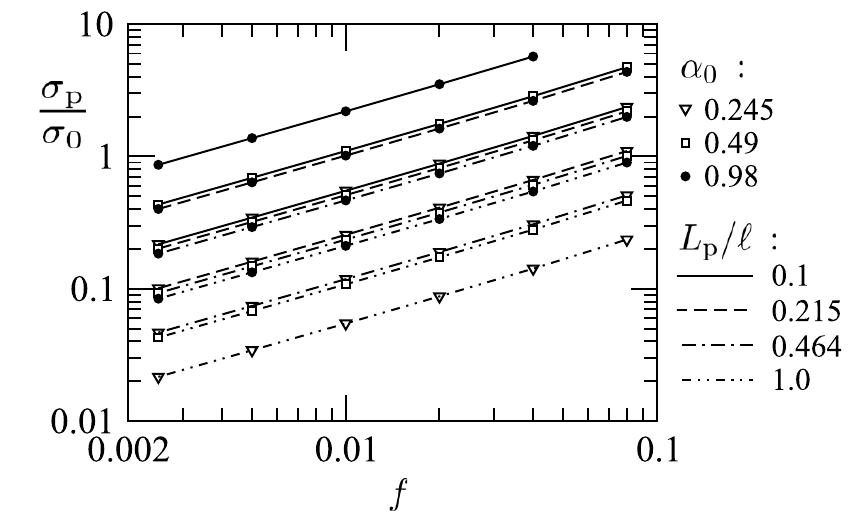} }
\caption{Influence of volume fraction of particles on the stress-strain response. (a) The macroscopic stress-strain response, and (b) increase in strength versus volume fraction for twelve combinations of parameters. Parameters used to generate the curves are indicated in the respective graph.}
\label{fig:fig_f}
\end{center}
\end{figure}

Figure \ref{fig:fig_f}a shows a set of stress-strain curves for $f$ in the range 0.025 to 0.8, generated with constant values of $L_{\rm p}/\ell = 0.464$ and $\alpha_0 = 0.49$. Strengthening increase with volume fraction and so does the post yield strain hardening $h_{\rm p}$. By a closer examination it was found that the post yield strain hardening is essentially linear with a slope approximately given by ${\rm d} h_{\rm p}/(G {\rm d}E_{\rm e}^{\rm p}) = 1.57f + 2.38f^2$, and was observed to be independent of $\alpha_0$ and $L_{\rm p}/\ell$. This result is similar to the result obtained from 2D axisymmetric analysis in \cite{Asgharzadeh2021}.

In Figure \ref{fig:fig_f}b,  $\sigma_{\rm p}/\sigma_0$ is plotted versus $f$ in a log-log diagram for twelve combinations of $\alpha_0$ and $L_{\rm p}/\ell$. The curves are essentially linear, and a careful inspection revealed that the curves have a slope equal to $2/3$ for sufficiently small values of $f$, which confirm Eq. \eref{eqn:Sp3Dcell_rconst}.

\subsection{Influence of variations in particle spacing, $C_{\rm v}^{\rm spacing}$}

\begin{figure}[htb!]
\begin{center}
    \includegraphics[width=0.60\textwidth]{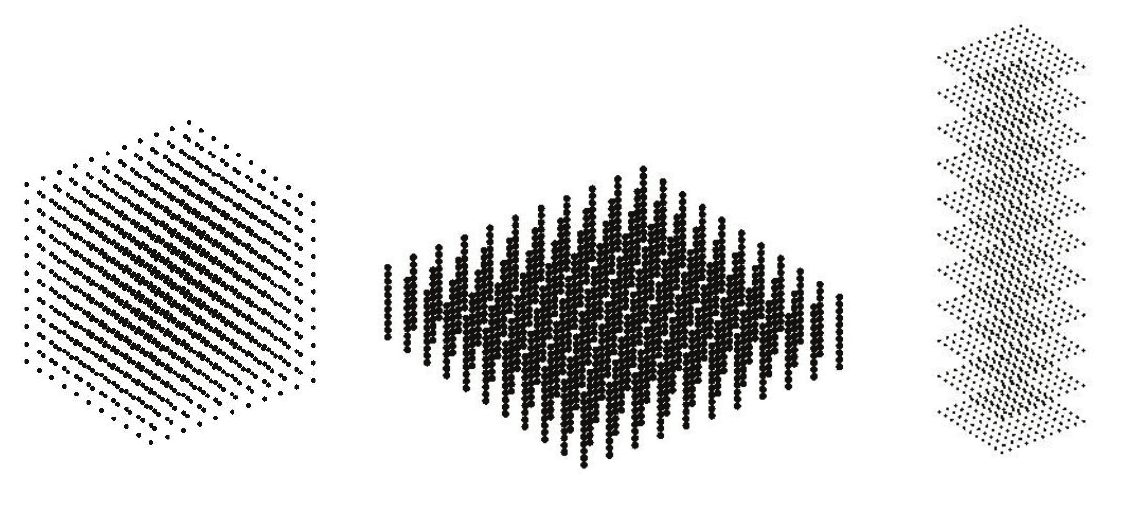}
\caption{Distributions of particles for extreme values of $C_{\rm v}^{\rm spacing}$. From left to right: $C/A = 1.0$, $C/A \ll 1.0$ (columns of particles), and $C/A \gg 1.0$ (planes of particles).}
\label{fig:C_v_distributions}
\end{center}
\end{figure}

\noindent The present unit cell model allows for an investigation of how an inhomogeneous particle distribution may affect strengthening. By varying the ratio between height and base of the unit cell ($C/A$), the particle spacing distribution will become inhomogeneous with $C_{\rm v}^{\rm spacing} > 0$. For $C/A$ sufficiently smaller than unity, the resulting distribution may be described as columns with tightly packed particles, whereas a $C/A$ ratio significantly larger than unity leads to planes with tightly packed particles, as illustrated in Figure \ref{fig:C_v_distributions}.

\begin{figure}[htb!]
\begin{center}
    \subfigure[]{ \includegraphics[scale=0.96]{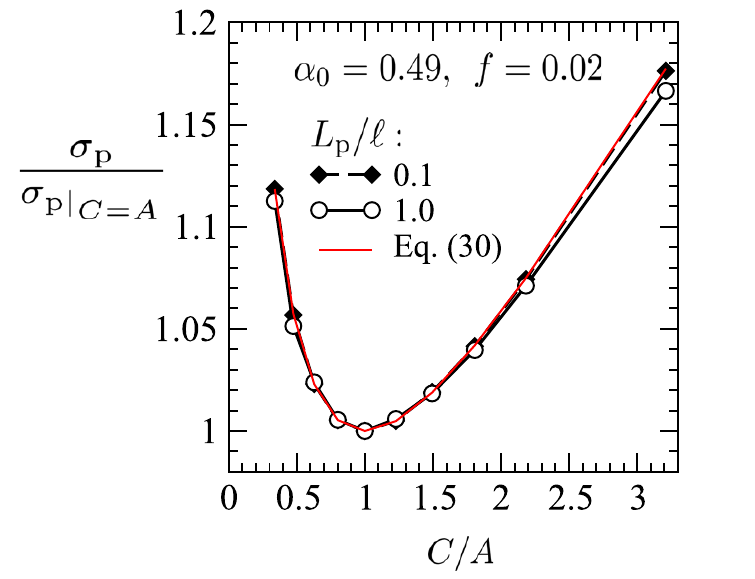} }
    \subfigure[]{ \includegraphics[scale=0.96]{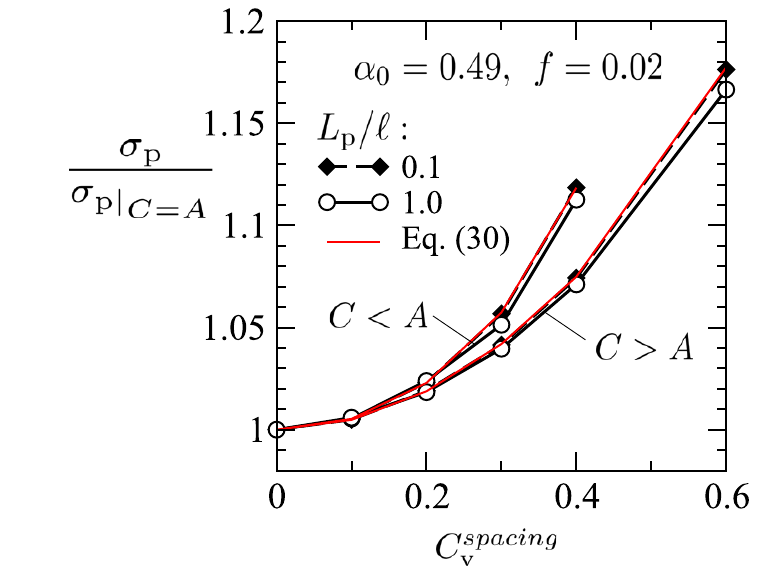} }
\caption{Influence of inhomogeneity in particle spacing distribution on the increase of yield strength. Relative strengthening is plotted versus $C/A$ in (a), and versus $C_{\rm v}^{\rm spacing}$ in (b). Black lines are for numerical results, while red lines are the predictions from of Eq. \eref{eqn:Sp3Dcell_rconst}.}
\label{fig:C_v_spacing_results}
\end{center}
\end{figure}

Two sets of curves with different average center-to-center particle spacing ($L_{\rm p}/\ell$) were generated with a constant volume fraction of particles $f = 0.02$ and $\alpha_0 = 0.49$. In one set particles are relatively small and closely spaced ($L_{\rm p}/\ell = 0.1$), and in the other set, particles are ten times larger and consequently much more sparsely spaced ($L_{\rm p}/\ell = 1$). The results are presented in Figure \ref{fig:C_v_spacing_results}a, where the increase in yield stress $\sigma_{\rm p}$ normalized by the value obtained for a homogeneous distribution ($C=A$) is plotted versus $C/A$ in the range 0.338 to 3.21. It can be observed that strengthening has a minimum for a homogeneous distribution and is larger for inhomogeneous distributions ($C \ne A$). Furthermore, both sets agree well with the prediction from Eq. \eref{eqn:Sp3Dcell_rconst}, included as a red solid curve.  The result presented in Fig. \ref{fig:C_v_spacing_results}a is replotted in Fig. \ref{fig:C_v_spacing_results}b, but as a function of the coefficient of variation ($C_{\rm v}^{\rm spacing}$) instead. As can be seen, $C_{\rm v}^{\rm spacing}$ is not an unambiguous parameter for characterizing strengthening effects when the particle spacing is inhomogeneous, because the two branches $C<A$ and $C>A$ do not collapse to give the same strengthening if evaluated at the same value of $C/A ~ (=\xi)$. This is in concord with the results obtained with the 2D axisymmetric model in \cite{Asgharzadeh2021}.

\subsection{Influence of variations in particle size, $C_{\rm v}^{\rm size}$}

\begin{figure}[htb!]
\begin{center}
    \includegraphics[scale=0.96]{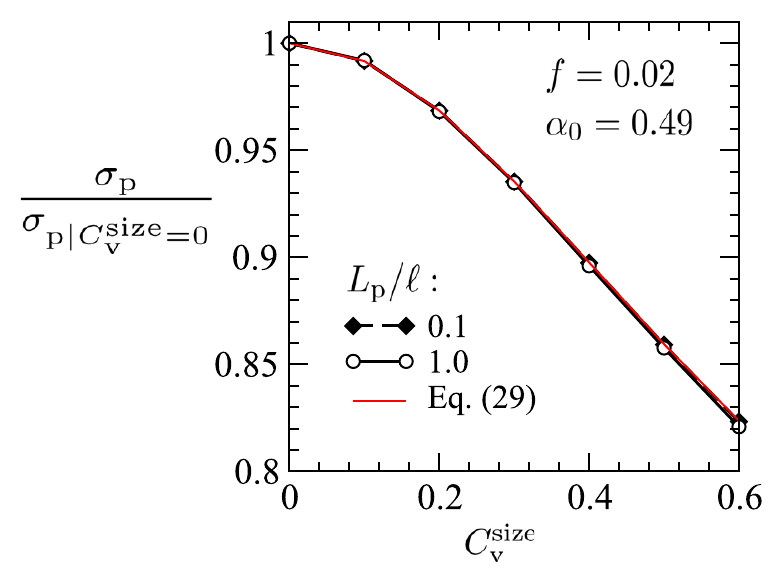}
\caption{Influence of inhomogeneity in particle size distribution on the increase of yield strength. Black lines are from numerical analyses, while red line is the prediction by Eq. \eref{eqn:Sp3Dcell_rvar}.}
\label{fig:C_v_Sizes}
\end{center}
\end{figure}

\noindent Again, two sets of curves with different average center-to-center particle spacing ($L_{\rm p}/\ell = 0.1, ~1$) were generated with a constant volume fraction of particles $f = 0.02$ and $\alpha_0 = 0.49$. Discrete particle size distributions were generated according to Eqs. \eref{eqn:pdf_D} - \eref{eqn:pdf_ri} where $f = 0.02$ was enforced by use of \eref{eqn:f_equation}. The result is presented in Figure \ref{fig:C_v_Sizes} as the increase in yield stress $\sigma_{\rm p}$ normalized by the value obtained for a homogeneous size distribution (all $r_{\rm i}$ equal) plotted versus $C_{\rm v}^{\rm size}$ in the range 0 to 0.6. As noted, a variance in particle size has a negative impact on strengthening. The analysis was repeated for a number of dissimilar positioning schemes of the eight particles of different size in the unit cell shown in Figure \ref{fig:Distribution_RVE}. However, the same result was obtained independent of the positioning. Finally, predictions from the proposed strengthening relation \eref{eqn:Sp3Dcell_rvar} fall on top of the numerical results as can be observed in Figure \ref{fig:C_v_Sizes}.

\subsection{Influence of particles with different strengths}
\label{the_two_examples}

\begin{figure}[htb!]
	\begin{center}
		\subfigure[]{ \includegraphics[scale=0.96]{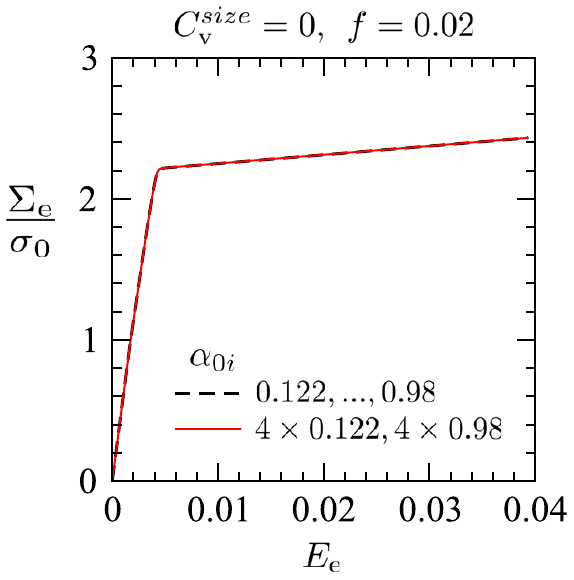} }
		\subfigure[]{ \includegraphics[scale=0.96]{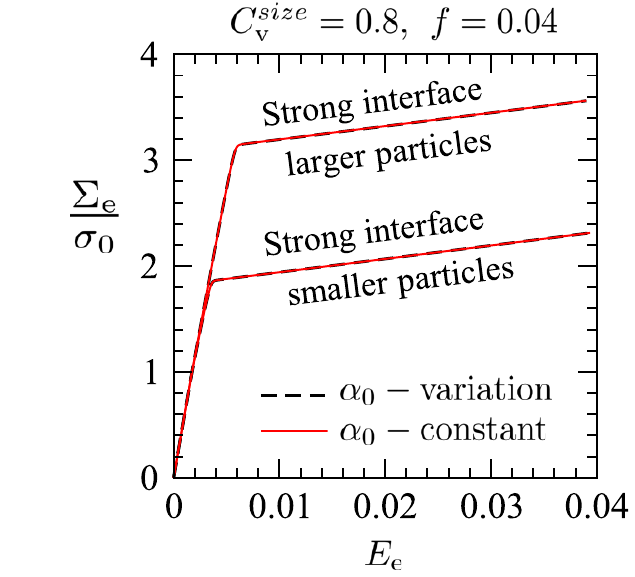} }
		\caption{Stress-strain curves for two examples of particles with different interface strength. (a) Equal sized particles: dashed black lines shows results for particles with $\alpha_0 = [0.1225, ..., 0.980]$; solid red lines show the result for a case where $\alpha_0 = 4 \times 0.1225, 4 \times 0.98$.  (b) Particles with a size distribution defined by $C_{\rm v}^{\rm size} = 0.8$, where the interface strength $\alpha_0 = [0.1225, ..., 0.98]$ either increases or decreases with particle size. As a reference two cases are included where equivalent values of $\alpha_0$ were chosen according to Eq. \eref{eqn:Sp3Dcell_rvar}.}
		\label{fig:the_two_examples}
	\end{center}
\end{figure}

\noindent Two examples will now be presented that illuminate effects of a varying interface strength. In the first example, a unit cell with $C_{\rm v}^{\rm spacing} = C_{\rm v}^{\rm size} = 0.0$ is considered. A volume fraction of $f = 0.02$ and a particle spacing of $L_{\rm p}/l = 0.1$ were chosen. In the first set, the $\alpha_0$ values $[0.1225, 0.245, 0.367, 0.490, 0.612, 0.735, 0.857, 0.980]$ were assigned to the eight particles. The solid black line in Figure \ref{fig:the_two_examples}a shows the stress-strain curve for this set. In a second set, $\alpha_0 = 0.1225$ was assigned to four particles, and $\alpha_0 = 0.98$ was assigned to the remaining four particles. According to \eref{eqn:Sp3Dcell_rconst} these two sets should result in identical stress-strain curves. Indeed, this is the case as seen by the red solid line falling on top of the dashed black line in Figure \ref{fig:the_two_examples}a.

In the second example, a distribution of particle sizes was considered with $C_{\rm v}^{\rm size} = 0.8$. A volume fraction of $f = 0.04$ and a particle spacing of $L_{\rm p}/l = 0.1$ were chosen. In the first set, eight different values $\alpha_0 = [0.1225, 0.245, 0.367, 0.490, 0.612, 0.735, 0.857, 0.980]$ were assigned to particles in increasing order of size, which means that $\alpha_0 = 0.1225$ was assigned to the smallest particle and $\alpha_0 = 0.980$ was assigned to the largest particle. In a second set, the same values of $\alpha_0$ were assigned to particles, but this time in a reversed order of particle size. The solid black lines in Figure \ref{fig:the_two_examples}b show the resulting stress-strain response of these two sets. Again, it is possible to choose two equivalent and constant $\alpha_0$ values that according to \eref{eqn:Sp3Dcell_rconst} would give the same macroscopic behaviour as obtained with the two sets above. For the first set, the equivalent value of $\alpha_0$ is $0.840$, and for the second set it becomes $0.262$. The results of these two equivalent sets are plotted as solid red lines in Figure \ref{fig:the_two_examples}b. As noted, the proposed strengthening expression is highly accurate also for these examples.

\begin{figure}[htb!]
	\begin{center}
		\centering
		\subfigure[]{ \includegraphics[scale=0.96]{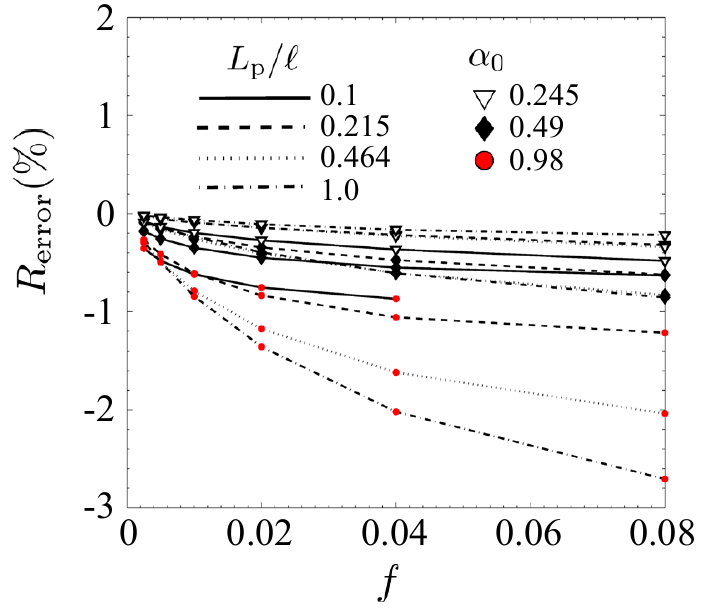} }\\
		\subfigure[]{ \includegraphics[scale=0.9]{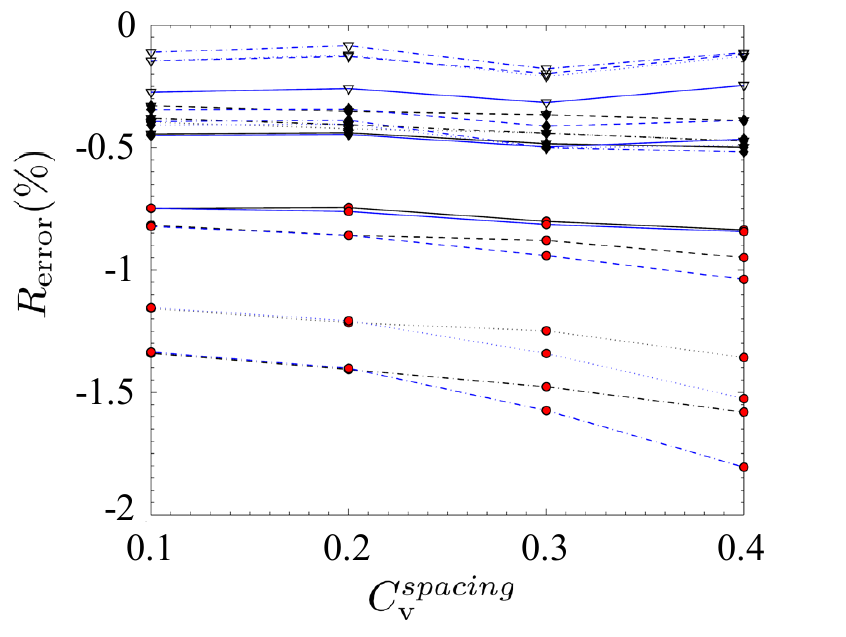} }
		\subfigure[]{ \includegraphics[scale=0.9]{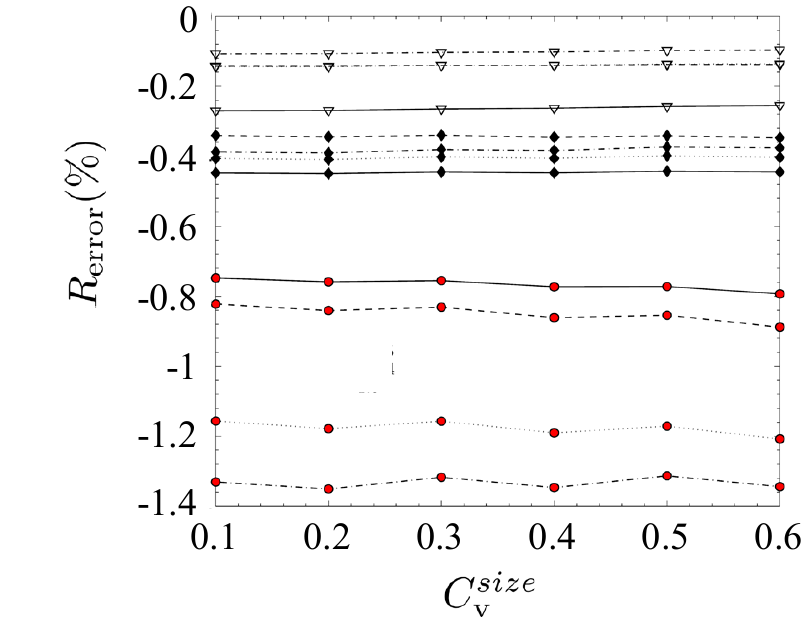} }
		\caption{Relative error of predicted strengthening using the general equation, compared to obtained results of numerical analysis. (a) Relative error versus $f$. (b) Relative error versus $C_{\rm v}^{\rm spacing}$ (blue lines are for $C/A < 1.0$ values). (c) Relative error versus $C_{\rm v}^{\rm size}$. }
		\label{fig:relative_errors}
	\end{center}
\end{figure}

\subsection{Relative error of the proposed strengthening relation}  \label{results_summary}

\noindent To summarize, all results presented in Figures \ref{fig:fig_f}, \ref{fig:C_v_spacing_results} and \ref{fig:C_v_Sizes} will now be used to give an overview of the accuracy of the proposed expression for strengthening \eref{eqn:Fs_function}, as adapted to the current 3D unit cell model in Eqs. \eref{eqn:Sp3Dcell_rvar} and \eref{eqn:Sp3Dcell_rconst}. As a measure of accuracy, the relative error in the macroscopic yield stress defined as $R_{\rm error} = \sigma_{\rm yFEM}/(\sigma_0 + \sigma_{\rm p}) - 1$ will be used ($\sigma_{\rm p}$ was calculated by use of \eref{eqn:Sp3Dcell_rvar} or \eref{eqn:Sp3Dcell_rconst}).
The results are shown in Figure \ref{fig:relative_errors}. In most cases the relative error is within 1 \%, except in a few cases where $L_{\rm p} / \ell$ approaches one.

\section{Concluding remarks}
\label{concluding_section}

\noindent The strengthening model  \eref{eqn:Fs_function} for a particle reinforced strain gradient plasticity material proposed by \cite{Asgharzadeh2021}, developed based on 2D axisymmetric finite element calculations, has been validated by use of a 3D unit cell model containing eight particles. For average center-to-center particle spacing $L_{\rm p}$ less than the material length parameter $\ell$ of the SGP material, the accuracy of the proposed relation appears to be within 1 \%.

An inhomogeneous spacing distribution of particles ($C \ne A$) is be beneficial for strengthening, as opposed to a variation in particle size, which causes a deterioration of strengthening as compared to a material with equal sized particles.

\section*{Acknowledgements}
This work was performed within the VINN Excellence Center Hero-m, financed by VINNOVA, the Swedish Governmental Agency for Innovation Systems, Swedish industry, and KTH Royal Institute of Technology.

\begin{appendices}
\renewcommand{\theequation}{\Alph{section}.\arabic{equation}}
\counterwithin*{equation}{section}

\section{Finite element discretization and implementation}
\label{appendA}

\subsection{Finite element discretization of the continuum}
Displacement and plastic strain fields in an element are given by the nodal DOFs defined in \eref{eqn:dofu} and \eref{eqn:dofep} and shape functions as
\begin{equation}
\left[ \begin{array}{c}
\vect{u} \\ 
\vecsym{\varepsilon}^{\rm p} 
\end{array} \right] = {\left[ \begin{array}{cc}
\mathds{N}_{\rm u} & \mathds{O} \\ 
\mathds{O} & \mathds{N}_{\rm p} 
\end{array} \right]} \left[ \begin{array}{c} \vect{d}_{\rm u} \\ \vect{d}_{\rm p} \end{array} \right] = \mathds{N}  \left[ \begin{array}{c} \vect{d}_{\rm u} \\ \vect{d}_{\rm p} \end{array} \right],
\end{equation}
where $\vect{u}^{T} = [u_{x} \quad u_{y} \quad u_{z}]$, and $(\vecsym{\varepsilon}^{\rm p})^{T} = [{\epsilon_{xx}^{{\rm p}}} \quad {\epsilon_{yy}^{{\rm p}}} \quad {\epsilon_{zz}^{{\rm p}}} \quad {\gamma_{xy}^{{\rm p}}} \quad {\gamma_{xz}^{{\rm p}}} \quad {\gamma_{yz}^{{\rm p}}}]$, $ \mathds{O} $ is a zero matrix of appropriate dimension,  and $\mathds{N}_{\rm u}$ and $\mathds{N}_{\rm p}$ are matrices containing shape functions for displacement and plastic strain degrees of freedom according to
\begin{equation}
\mathds{N}_{\rm u} = \left[ \begin{array}{cccccccccc}
N_{\rm u}^{1} & 0 & 0 & N_{\rm u}^{2} & 0 & 0 & \ldots & N_{\rm u}^{10} & 0 & 0 \\
0 & N_{\rm u}^{1} & 0 & 0 & N_{\rm u}^{2} & 0 & \ldots & 0 & N_{\rm u}^{10} & 0 \\
0 & 0 & N_{\rm u}^{1} & 0 & 0 & N_{\rm u}^{2} & \ldots & 0 & 0 & N_{\rm u}^{10}
\end{array} \right]_{3 \times 30},
\end{equation}

\begin{equation} \label{eqn:shapefcnNp}
\mathds{N}_{\rm p} = \left[ \begin{array}{ccccccccccc}
N_{\rm p}^{1} & 0 & 0 & 0 & 0 & \ldots & N_{\rm p}^{4} & 0 & 0 & 0 & 0 \\
0 & N_{\rm p}^{1} & 0 & 0 & 0 & \ldots & 0 & N_{\rm p}^{4} & 0 & 0 & 0 \\
-N_{\rm p}^{1} & -N_{\rm p}^{1} & 0 & 0 & 0 & \ldots & -N_{\rm p}^{4} & -N_{\rm p}^{4} & 0 & 0 & 0 \\
0 & 0 & N_{\rm p}^{1} & 0 & 0 & \ldots & 0 & 0 & N_{\rm p}^{4} & 0 & 0 \\
0 & 0 & 0 & N_{\rm p}^{1} & 0 & \ldots & 0 & 0 & 0 & N_{\rm p}^{4} & 0 \\
0 & 0 & 0 & 0 & N_{\rm p}^{1} & \ldots & 0 & 0 & 0 & 0 & N_{\rm p}^{4}
\end{array} \right]_{6 \times 20} .
\end{equation}
Here, complete quadratic interpolation functions are used for displacements and linear interpolation functions are used for plastic strains. Note that plastic incompressibility ($\varepsilon_{zz}^{\rm p} = -(\varepsilon_{xx}^{\rm p} + \varepsilon_{yy}^{\rm p})$) is directly enforced by the third row in \eref{eqn:shapefcnNp}. The relevant strain fields and gradients thereof needed for calculation of stresses and the consistent tangent stiffness matrix in an element are
\begin{align}
(\vecsym{\varepsilon}^{\rm e})^{\rm T} & = [ \varepsilon^{\rm e}_{xx} ~ \varepsilon^{\rm e}_{yy} ~ \varepsilon^{\rm e}_{zz} ~ \gamma^{\rm e}_{xy} ~ \gamma^{\rm e}_{xz} ~ \gamma^{\rm e}_{xz}], \notag \\
(\vecsym{\varepsilon}^{\rm p})^{\rm T} & = [ \varepsilon^{\rm p}_{xx} ~ \varepsilon^{\rm p}_{yy} ~ \varepsilon^{\rm p}_{zz} ~ \gamma^{\rm p}_{xy} ~ \gamma^{\rm p}_{xz} ~ \gamma
^{\rm p}_{yz}], \\
(\nabla\vecsym{\varepsilon}^{\rm p})^{\rm T} & = [ \varepsilon^{\rm p}_{xx,x} ~ \varepsilon^{\rm p}_{yy,x} ~ \varepsilon^{\rm p}_{zz,x} ~ \gamma^{\rm p}_{xy,x} ~ \gamma^{\rm p}_{xz,x} ~ \gamma^{\rm p}_{yz,x} ~ \varepsilon^{\rm p}_{xx,y} ~ \varepsilon^{\rm p}_{yy,y} ~ \varepsilon^{\rm p}_{zz,y} ~ \gamma^{\rm p}_{xy,y} ~ \gamma^{\rm p}_{xz,y} ~ \gamma^{\rm p}_{yz,y} ] \notag  \\ & \quad \quad \varepsilon^{\rm p}_{xx,z} ~ \varepsilon^{\rm p}_{yy,z} ~ \varepsilon^{\rm p}_{zz,z} ~ \gamma^{\rm p}_{xy,z} ~ \gamma^{\rm p}_{xz,z} ~ \gamma^{\rm p}_{yz,z} ] \notag.
\end{align}
These are evaluated by use of the nodal DOFs as
\begin{equation} \label{eqn:element_strains}
\left[ \begin{array}{c} \vecsym{\varepsilon}^{\rm e} \\ \vecsym{\varepsilon}^{\rm p} \\ \nabla\vecsym{\varepsilon}^{\rm p} \end{array} \right] = {\left[ \begin{array}{cc} \mathds{B}_{\rm u} & -\mathds{N}_{\rm p} \\ \mathds{O} & \mathds{N}_{\rm p} \\ \mathds{O} & \mathds{B}_{\rm p} \end{array} \right]} \left[ \begin{array}{c} \vect{d}_{\rm u}  \\ \vect{d}_{\rm p} \end{array} \right] = \mathds{B} \left[ \begin{array}{c} \vect{d}_{\rm u}  \\ \vect{d}_{\rm p} \end{array} \right].
\end{equation}
In \eref{eqn:element_strains}, matrix $\mathds{B}$ is of size $30 \times 50$ and contains gradient operators. Complete quadratic polynomials are used for the interpolation of spatial coordinates for a material point in an element as
\begin{equation}
	x_{i}(\xi_{k}) = \sum_{I=1}^{10} N_{\rm u}^{I}(\xi_{k})x_{i}^{I}, \qquad i = 1,2,3
\end{equation}
The gradient operators involve spatial derivatives of shape functions derived in a standard manner according to
\begin{align}
		(B_{\rm u}^{I})_{j} & = \frac{\partial N_{\rm u}^{I}}{\partial x_{j}} = J_{ij}^{-1} \frac{\partial N_{\rm u}^{I}}{\partial \xi_{i}}, \notag \\
		(B_{\rm p}^{I})_{j} & = \frac{\partial N_{\rm p}^{I}}{\partial x_{j}} = J_{ij}^{-1}\frac{\partial N_{\rm p}^{I}}{\partial \xi_{i}},
\end{align}
with
\begin{equation}
	J_{ij} =  \sum_{I=1}^{10} \frac{\partial N_{\rm u}^{I}}{\partial \xi_{i}} x_{j}^{I}.
\end{equation}
Note that the same Jacobian ($J_{ij}$) is used in both operators. Hence, only the description of displacements are isoparametric. The sub-matrices in \eref{eqn:element_strains} becomes explicitly
\begin{equation}
	\mathds{B}_{\rm u} = \left[ \begin{array}{ccccccc}
		(B_{\rm u}^{1})_{1} & 0 & 0 & \ldots & (B_{\rm u}^{10})_{1} & 0 & 0 \\
		0 & (B_{\rm u}^{1})_{2} & 0 & \ldots & 0 & (B_{\rm u}^{10})_{2} & 0 \\
		0 & 0 & (B_{\rm u}^{1})_{3} & \ldots & 0 & 0 & (B_{\rm u}^{10})_{3} \\
		(B_{\rm u}^{1})_{2} & (B_{\rm u}^{1})_{1} & 0 & \ldots & (B_{\rm u}^{10})_{2} & (B_{\rm u}^{10})_{1} & 0 \\
		(B_{\rm u}^{1})_{3} & 0 & (B_{\rm u}^{1})_{1} & \ldots & (B_{\rm u}^{10})_{3} & 0 & (B_{\rm u}^{10})_{1} \\
		0 & (B_{\rm u}^{1})_{3} & (B_{\rm u}^{1})_{2} & \ldots & 0 & (B_{\rm u}^{10})_{3} & (B_{\rm u}^{10})_{2}
	\end{array} \right]_{6 \times 30}
\end{equation}
and
\begin{equation}
	\mathds{B}_{\rm p} = \left[ \mathds{B}_{\rm p}^{1} \quad \mathds{B}_{\rm p}^{2} \quad \ldots \mathds{B}_{\rm p}^{4} \right]_{18 \times 20},
\end{equation}
where
\begin{equation}
	\mathds{B}_{\rm p}^{I} = \left[ \begin{array}{ccccc}
		(B_{\rm p}^{I})_{1} & 0 & 0 & 0 & 0 \\
		0 & (B_{\rm p}^{I})_{1} & 0 & 0 & 0 \\
		-(B_{\rm p}^{I})_{1} & -(B_{\rm p}^{I})_{1} & 0 & 0 & 0 \\
		0 & 0 & (B_{\rm p}^{I})_{1} & 0 & 0 \\
		0 & 0 & 0 & (B_{\rm p}^{I})_{1} & 0 \\
		0 & 0 & 0 & 0 & (B_{\rm p}^{I})_{1} \\
		\vdots & \vdots & \vdots & \vdots & \vdots \\
		(B_{\rm p}^{I})_{3} & 0 & 0 & 0 & 0 \\
		0 & (B_{\rm p}^{I})_{3} & 0 & 0 & 0 \\
		-(B_{\rm p}^{I})_{3} & -(B_{\rm p}^{I})_{3} & 0 & 0 & 0 \\
		0 & 0 & (B_{\rm p}^{I})_{3} & 0 & 0 \\
		0 & 0 & 0 & (B_{\rm p}^{I})_{3} & 0 \\
		0 & 0 & 0 & 0 & (B_{\rm p}^{I})_{3}
	\end{array} \right]_{18 \times 5}.
\end{equation}
The stress-like quantities are collected in a vector
\begin{equation}
\vect{s} = \left[ \begin{array}{c} \vecsym{\sigma} \\ \vect{q} \\ \vect{m} \end{array} \right]
\end{equation}
where
\begin{align}
	\vecsym{\sigma}^{\rm T} & = [ \sigma_{xx}  ~ \sigma_{yy}  ~ \sigma_{zz}  ~ \sigma_{xy} \quad \sigma_{xz}  ~ \sigma_{yz} ], \notag \\
	\vect{q}^{\rm T} & = [ q_{xx}  ~ q_{yy}  ~ q_{zz}  ~ q_{xy}  ~ q_{xz}  ~ q_{yz} ], \\
	\vect{m}^{\rm T} & = [ m_{xxx} ~ m_{yyx} ~ m_{zzx} ~ m_{xyx} ~ m_{xzx} ~ m_{yzx} ~ m_{xxy} ~ \ldots ~ m_{yzz} ].  \notag
\end{align}
The traction vector components are $ T_{i} = \sigma_{ij}n_{j} $ for the standard tractions and $ M_{ij} = m_{ijk}n_{k} $ for the moment tractions. These components are collected in vector as
\begin{equation} \label{eqn:fem_tractions}
\vect{t} = \left[ \begin{array}{c} \vect{T} \\ \vect{M} \end{array} \right]  = \left[ \begin{array}{c} \sigma_{xx}n_{x} + \sigma_{xy}n_{y} + \sigma_{xz}n_{z} \\ \sigma_{xy}n_{x} + \sigma_{yy}n_{y} + \sigma_{yz}n_{z} \\ \sigma_{xz}n_{x} + \sigma_{yz}n_{y} + \sigma_{zz}n_{z} \\ m_{xxx}n_{x} + m_{xxy}n_{y} + m_{xxz}n_{z} \\ m_{yyx}n_{x} + m_{yyy}n_{y} + m_{yyz}n_{z} \\ m_{zzx}n_{x} + m_{zzy}n_{y} + m_{zzz}n_{z} \\ m_{xyx}n_{x} + m_{xyy}n_{y} + m_{xyz}n_{z} \\ m_{xzx}n_{x} + m_{xzy}n_{y} + m_{xzz}n_{z} \\ m_{yzx}n_{x} + m_{yzy}n_{y} + m_{yzz}n_{z} \end{array} \right].
\end{equation}

\subsection{Finite element discretization of the interface}

We distinguish between two sides of an interface element: the particle side (1) and the matrix side (2). Since particles are assumed to be elastic, plastic strain field is present only on side (2). The interpolation of this field using nodal values is done through
\begin{equation}
\vecsym{\varepsilon}_{\Gamma} = \mtrx{N}_{\Gamma} \vect{d}_{\rm p}^{\rm (2)},
\end{equation}
where $ \mtrx{N}_{\Gamma} $ has the same structure as the corresponding matrix for bulk elements, but of dimensions $6 \times 15$ to comply with the interface element. The moment traction vector that is work conjugate to $\vecsym{\varepsilon}_{\Gamma}$ is denoted  $\vect{M}_{\Gamma}$ and corresponds to the lower sub-matrix in \eref{eqn:fem_tractions}.

\subsection{Enforcing equilibrium}
The nonlinear problem is solved incrementally by reducing the out-of-balance (residual) forces to zero. The residual force vector is obtained by discretization of the force balance equation derived from the virtual work relation \eref{eqn:VirtualWork} over the entire domain as
\begin{equation}\label{eqn:FEequilibrium}
\vect{r} = \vect{f}_{V} + \vect{f}_{S^{\Gamma}} - \vect{f}_{S^{\rm ext}} =  \int_V \mathds{B}^{\rm T} \vect{s}~{\rm d}V + \int_{S^{\Gamma}} \mtrx{N}_{\Gamma}^{\rm T}\vect{M}_{\Gamma}{\rm d}S - \int_{S^{\rm ext}} \mathds{N}^{\rm T} \vect{t}~{\rm d}S.
\end{equation}

\noindent An Euler backward algorithm using with a full Newton iteration scheme is used to progress the solution a load increment. Denoting an iteration by superscript $ k $, linearization of \eref{eqn:FEequilibrium} by Taylor expansion of increment $k + 1$ gives
\begin{equation}\label{eqn:FEequLinearized}
\vect{r}^{k+1} = \vect{r}^{k} + \frac{\partial \vect{r}^{k}}{\partial \mathbf{d}} \Delta\mathbf{d}^{k+1}.
\end{equation}

\noindent As external forces are constant over a load increment and independent of $ \mathbf{d} $, the partial derivative in \eref{eqn:FEequLinearized} is calculated according to
\begin{equation}\label{eqn:partialderivforKtan}
\frac{\partial \vect{r}^{k}}{\partial \mathbf{d}} = \int_V \mathds{B}^{\rm T} \frac{\partial \mathbf{s}^{k}}{\partial \boldsymbol{\epsilon}} \mathds{B}~{\rm d}V + \int_{S^{\Gamma}} \mathds{N}_{\Gamma}^{\rm T}\frac{\partial \mathbf{M}_{\Gamma}^{k}}{\partial \vecsym{\varepsilon}_{\Gamma}}\mathds{N}_{\Gamma} ~{\rm d}S,
\end{equation}
where the partial derivatives can be found from the constitutive relationships and use of the Euler-backward method and expressed by the consistent material point stiffness matrices $ \mathds{D}_V $ (continuum) and $\mathds{D}_{\Gamma}$ (interface) such that
\begin{equation}
\int_V \mathds{B}^{\rm T} \mathds{D}_V \mathds{B}~{\rm d}V + \int_{S^{\Gamma}} \mathds{N}_{\Gamma}^{\rm T} \mathds{D}_{\Gamma} \mathds{N}_{\Gamma} ~{\rm d}S = \mathds{K}_{\rm tan},
\end{equation}
which defines the tangent stiffness $ \mathds{K}_{\rm tan} $. Letting $ \vect{r}^{k+1} \rightarrow \vect{0} $ give the increment
\begin{equation}
\Delta \mathbf{d}^{k+1} = - (\mathds{K}_{\rm tan})^{-1} \mathbf{r}^{k}.
\end{equation}

\end{appendices}

\bibliography{ReferencesToB}

\begin{thebibliography}{28}
\expandafter\ifx\csname natexlab\endcsname\relax\def\natexlab#1{#1}\fi
\expandafter\ifx\csname url\endcsname\relax
  \def\url#1{\texttt{#1}}\fi
\expandafter\ifx\csname urlprefix\endcsname\relax\def\urlprefix{URL }\fi
\providecommand{\eprint}[2][]{\url{#2}}
\providecommand{\bibinfo}[2]{#2}
\ifx\xfnm\relax \def\xfnm[#1]{\unskip,\space#1}\fi
\bibitem[{Anand et~al.(2005)Anand, Gurtin, Lele and Gething}]{Anand05}
\bibinfo{author}{Anand, L.}, \bibinfo{author}{Gurtin, M.E.},
  \bibinfo{author}{Lele, S.P.}, \bibinfo{author}{Gething, C.},
  \bibinfo{year}{2005}.
\newblock \bibinfo{title}{A one-dimensional theory of strain-gradient
  plasticity: Formulation, analysis, numerical results}.
\newblock \bibinfo{journal}{Journal of the Mechanics and Physics of Solids}
  \bibinfo{volume}{53}, \bibinfo{pages}{1789--1826}.
\bibitem[{Ardell(1985)}]{Ardell85}
\bibinfo{author}{Ardell, A.J.}, \bibinfo{year}{1985}.
\newblock \bibinfo{title}{Precipitation hardening}.
\newblock \bibinfo{journal}{Metallurgical Transactions A} \bibinfo{volume}{16},
  \bibinfo{pages}{2131--2165}.
\bibitem[{Asgharzadeh and Faleskog(2021)}]{Asgharzadeh2021a}
\bibinfo{author}{Asgharzadeh, M.}, \bibinfo{author}{Faleskog, J.},
  \bibinfo{year}{2021}.
\newblock \bibinfo{title}{A strengthening model of particle-matrix interaction
  based on an axisymmetric strain gradient plasticity analysis}.
\newblock \bibinfo{note}{ArXiv preprint arXiv:2106.08432 [cond-mat.mtrl-sci]}.
\bibitem[{Ashby(1969)}]{ashby1969}
\bibinfo{author}{Ashby, M.}, \bibinfo{year}{1969}.
\newblock \bibinfo{title}{On the orowan stress}, in: \bibinfo{editor}{Argon,
  A.} (Ed.), \bibinfo{booktitle}{Physics of strength and plasticity}.
  \bibinfo{publisher}{MIT Press, Cambridge, MA}, pp. \bibinfo{pages}{113--131}.
\bibitem[{Ashby(1970)}]{Ashby70}
\bibinfo{author}{Ashby, M.F.}, \bibinfo{year}{1970}.
\newblock \bibinfo{title}{Deformation of plastically non-homogeneous
  materials}.
\newblock \bibinfo{journal}{Philosophical Magazine} \bibinfo{volume}{21},
  \bibinfo{pages}{399--424}.
\bibitem[{Azizi et~al.(2014)Azizi, Niordson and Legarth}]{Azizi14}
\bibinfo{author}{Azizi, R.}, \bibinfo{author}{Niordson, C.F.},
  \bibinfo{author}{Legarth, B.N.}, \bibinfo{year}{2014}.
\newblock \bibinfo{title}{On the homogenization of metal matrix composites
  using strain gradient plasticity}.
\newblock \bibinfo{journal}{Acta Mechanica Sinica} \bibinfo{volume}{30},
  \bibinfo{pages}{175--190}.
\bibitem[{Dahlberg and Faleskog(2013a)}]{Dahlberg2013b}
\bibinfo{author}{Dahlberg, C.F.}, \bibinfo{author}{Faleskog, J.},
  \bibinfo{year}{2013}a.
\newblock \bibinfo{title}{An improved strain gradient plasticity formulation
  with energetic interfaces: theory and a fully implicit finite element
  formulation}.
\newblock \bibinfo{journal}{Computational Mechanics} \bibinfo{volume}{51},
  \bibinfo{pages}{641--659}.
\bibitem[{Dahlberg et~al.(2013)Dahlberg, Faleskog, Niordson and
  Legarth}]{Dahlberg13b}
\bibinfo{author}{Dahlberg, C.F.}, \bibinfo{author}{Faleskog, J.},
  \bibinfo{author}{Niordson, C.F.}, \bibinfo{author}{Legarth, B.N.},
  \bibinfo{year}{2013}.
\newblock \bibinfo{title}{A deformation mechanism map for polycrystals modeled
  using strain gradient plasticity and interfaces that slide and separate}.
\newblock \bibinfo{journal}{International Journal of Plasticity}
  \bibinfo{volume}{43}, \bibinfo{pages}{177--195}.
\bibitem[{Dahlberg and Faleskog(2013b)}]{Dahlberg13a}
\bibinfo{author}{Dahlberg, C.F.O.}, \bibinfo{author}{Faleskog, J.},
  \bibinfo{year}{2013}b.
\newblock \bibinfo{title}{An improved strain gradient plasticity formulation
  with energetic interfaces: theory and a fully implicit finite element
  formulation}.
\newblock \bibinfo{journal}{Computational Mechanics} \bibinfo{volume}{51},
  \bibinfo{pages}{641--659}.
\bibitem[{Deschamps and Brechet(1999)}]{Deschamps99}
\bibinfo{author}{Deschamps, A.}, \bibinfo{author}{Brechet, Y.},
  \bibinfo{year}{1999}.
\newblock \bibinfo{title}{Influence of predeformation and ageing of an al-zn-mg
  alloy-ii. modeling of precipitation kinetics and yield stress}.
\newblock \bibinfo{journal}{Acta Materialia} \bibinfo{volume}{47},
  \bibinfo{pages}{293--305}.
\bibitem[{Espinosa et~al.(2005)Espinosa, Berbenni, Panico and
  Schwarz}]{Espinosa2005}
\bibinfo{author}{Espinosa, H.D.}, \bibinfo{author}{Berbenni, S.},
  \bibinfo{author}{Panico, M.}, \bibinfo{author}{Schwarz, K.W.},
  \bibinfo{year}{2005}.
\newblock \bibinfo{title}{An interpretation of size-scale plasticity in
  geometrically confined systems}.
\newblock \bibinfo{journal}{Proceedings of the National Academy of Sciences of
  the United States of America} \bibinfo{volume}{102},
  \bibinfo{pages}{16933--16938}.
\bibitem[{Fleck and Hutchinson(1997)}]{Fleck97}
\bibinfo{author}{Fleck, N.A.}, \bibinfo{author}{Hutchinson, J.W.},
  \bibinfo{year}{1997}.
\newblock \bibinfo{title}{A phenomenological theory for strain gradient effects
  in plasticity}.
\newblock \bibinfo{journal}{Advances in Applied Mechanics}
  \bibinfo{volume}{33}, \bibinfo{pages}{295--361}.
\bibitem[{Foreman and Makin(1966)}]{Foreman66}
\bibinfo{author}{Foreman, A.J.E.}, \bibinfo{author}{Makin, M.J.},
  \bibinfo{year}{1966}.
\newblock \bibinfo{title}{Dislocation movement through random arrays of
  obstacles}.
\newblock \bibinfo{journal}{Philosophical Magazine} \bibinfo{volume}{14},
  \bibinfo{pages}{911--924}.
\bibitem[{Fredriksson and Gudmundson(2007)}]{Fredriksson07}
\bibinfo{author}{Fredriksson, P.}, \bibinfo{author}{Gudmundson, P.},
  \bibinfo{year}{2007}.
\newblock \bibinfo{title}{Modelling of the interface between a thin film and a
  substrate within a strain gradient plasticity framework}.
\newblock \bibinfo{journal}{Journal of the Mechanics and Physics of Solids}
  \bibinfo{volume}{55}, \bibinfo{pages}{939--955}.
\bibitem[{Fredriksson et~al.(2009)Fredriksson, Gudmundson and
  Mikkelsen}]{Fredriksson09}
\bibinfo{author}{Fredriksson, P.}, \bibinfo{author}{Gudmundson, P.},
  \bibinfo{author}{Mikkelsen, L.P.}, \bibinfo{year}{2009}.
\newblock \bibinfo{title}{Finite element implementation and numerical issues of
  strain gradient plasticity with application to metal matrix composites}.
\newblock \bibinfo{journal}{International Journal of Solids and Structures}
  \bibinfo{volume}{46}, \bibinfo{pages}{3977--3987}.
\bibitem[{Friedel(1964)}]{friedel64}
\bibinfo{author}{Friedel, J.}, \bibinfo{year}{1964}.
\newblock \bibinfo{title}{Dislocations: International Series of Monographs on
  Solid State Physics}.
\newblock International series of monographs on solid state physics,
  \bibinfo{publisher}{Elsevier Science}.
\bibitem[{Gao et~al.(1999)Gao, Huang, Nix and Hutchinson}]{Gao99}
\bibinfo{author}{Gao, H.}, \bibinfo{author}{Huang, Y.}, \bibinfo{author}{Nix,
  W.D.}, \bibinfo{author}{Hutchinson, J.W.}, \bibinfo{year}{1999}.
\newblock \bibinfo{title}{Mechanism based strain gradient plasticity--i.
  theory}.
\newblock \bibinfo{journal}{Journal of the Mechanics and Physics of Solids}
  \bibinfo{volume}{47}, \bibinfo{pages}{1239--1263}.
\bibitem[{Van~der Giessen and Needleman(1995)}]{Giessen95}
\bibinfo{author}{Van~der Giessen, E.}, \bibinfo{author}{Needleman, A.},
  \bibinfo{year}{1995}.
\newblock \bibinfo{title}{Discrete dislocation plasticity: A simple planar
  model}.
\newblock \bibinfo{journal}{Modelling and Simulation in Materials Science and
  Engineering} \bibinfo{volume}{3}, \bibinfo{pages}{689--735}.
\bibitem[{Gladman(1999)}]{Gladman1999}
\bibinfo{author}{Gladman, T.}, \bibinfo{year}{1999}.
\newblock \bibinfo{title}{Precipitation hardening in metals}.
\newblock \bibinfo{journal}{Materials Science and Technology}
  \bibinfo{volume}{15}, \bibinfo{pages}{30--36}.
\bibitem[{Gudmundson(2004)}]{Gudmundson04}
\bibinfo{author}{Gudmundson, P.}, \bibinfo{year}{2004}.
\newblock \bibinfo{title}{A unified treatment of strain gradient plasticity}.
\newblock \bibinfo{journal}{Journal of the Mechanics and Physics of Solids}
  \bibinfo{volume}{52}, \bibinfo{pages}{1379--1406}.
\bibitem[{Hu and Curtin(2021)}]{hu2021modeling}
\bibinfo{author}{Hu, Y.}, \bibinfo{author}{Curtin, W.}, \bibinfo{year}{2021}.
\newblock \bibinfo{title}{Modeling peak-aged precipitate strengthening in
  al--mg--si alloys}.
\newblock \bibinfo{journal}{Journal of the Mechanics and Physics of Solids}
  \bibinfo{volume}{151}, \bibinfo{pages}{1--19}.
\bibitem[{Hutchinson(2000)}]{Hutchinson2000}
\bibinfo{author}{Hutchinson, J.W.}, \bibinfo{year}{2000}.
\newblock \bibinfo{title}{Plasticity at the micron scale}.
\newblock \bibinfo{journal}{International Journal of Solids and Structures}
  \bibinfo{volume}{37}, \bibinfo{pages}{225--238}.
\bibitem[{Lubarda(2016)}]{Lubarda16}
\bibinfo{author}{Lubarda, V.A.}, \bibinfo{year}{2016}.
\newblock \bibinfo{title}{On the recoverable and dissipative parts of higher
  order stresses in strain gradient plasticity}.
\newblock \bibinfo{journal}{International Journal of Plasticity}
  \bibinfo{volume}{78}, \bibinfo{pages}{26--43}.
\bibitem[{Monnet(2015)}]{monnet2015multiscale}
\bibinfo{author}{Monnet, G.}, \bibinfo{year}{2015}.
\newblock \bibinfo{title}{Multiscale modeling of precipitation hardening:
  Application to the fe--cr alloys}.
\newblock \bibinfo{journal}{Acta Materialia} \bibinfo{volume}{95},
  \bibinfo{pages}{302--311}.
\bibitem[{Orowan(1948)}]{Orowan48}
\bibinfo{author}{Orowan, E.}, \bibinfo{year}{1948}.
\newblock \bibinfo{title}{Discussion on internal stresses}, in:
  \bibinfo{booktitle}{Symposium Internal stress in metals and alloys}.
  \bibinfo{publisher}{Institute of Metals}, pp. \bibinfo{pages}{451--453}.
\bibitem[{Reppich(1993)}]{Reppich93}
\bibinfo{author}{Reppich, B.}, \bibinfo{year}{1993}.
\newblock \bibinfo{title}{Particle strengthening}.
\newblock \bibinfo{journal}{Materials science and technology}
  \bibinfo{volume}{6}, \bibinfo{pages}{311--357}.
\bibitem[{Voyiadjis and Song(2019)}]{voyiadjis2019strain}
\bibinfo{author}{Voyiadjis, G.Z.}, \bibinfo{author}{Song, Y.},
  \bibinfo{year}{2019}.
\newblock \bibinfo{title}{Strain gradient continuum plasticity theories:
  theoretical, numerical and experimental investigations}.
\newblock \bibinfo{journal}{International Journal of Plasticity}
  \bibinfo{volume}{121}, \bibinfo{pages}{21--75}.
\bibitem[{Xue et~al.(2002)Xue, Huang and Li}]{Xue02}
\bibinfo{author}{Xue, Z.}, \bibinfo{author}{Huang, Y.}, \bibinfo{author}{Li,
  M.}, \bibinfo{year}{2002}.
\newblock \bibinfo{title}{Particle size effect in metallic materials: a study
  by the theory of mechanism-based strain gradient plasticity}.
\newblock \bibinfo{journal}{Acta Materialia} \bibinfo{volume}{50},
  \bibinfo{pages}{149--160}.

\end{thebibliography}

\end{document}